\documentclass[12pt]{article}
\usepackage{fullpage}
\usepackage{amsmath}
\usepackage{amsfonts}
\usepackage{latexsym}
\usepackage{color}
\begin{document}


\definecolor{light}{gray}{.85}
\newcommand{\hilite}[1]{\noindent\colorbox{light}{\parbox{\linewidth}{\addtolength{\parindent}{0.15in}\indent #1}}}

\newcommand{\mytheoremcounter}{section}
\newcommand{\cheapHack}{ }
\newcommand{\Paccept}[1]{{\Pr[#1 = \mathrm{accept}]}}
\newcommand{\Maj}{\mathit{Maj}}
\newcommand{\Nand}{\bar{\wedge}}
\newcommand{\Nor}{\bar{\vee}}
\newcommand{\myem}[1]{{\bf #1}}
\newcommand{\reccr}[1]{\overline{\mathrm{cr}}(#1)}
\newcommand{\regcr}[1]{\mathrm{cr}(#1)}
\newcommand{\NLclass}{\mathbf{NL}}
\newcommand{\NCclass}{\mathbf{NC}}
\newcommand{\coNLclass}{\mathbf{co\!-\!NL}}
\newcommand{\Lclass}{\mathbf{L}}
\newcommand{\Pclass}{\mathbf{P}}
\newcommand{\NPclass}{\mathbf{NP}}
\newcommand{\coNPclass}{\mathbf{co\!-\!NP}}
\newcommand{\PPclass}{\mathbf{PP}}
\newcommand{\BPPclass}{\mathbf{BPP}}
\newcommand{\ZPPclass}{\mathbf{ZPP}}
\newcommand{\RPclass}{\mathbf{RP}}
\newcommand{\coRPclass}{\mathbf{co\!-\!RP}}
\newcommand{\SigmaP}[1]{\mathbf{\Sigma_{#1}P}}
\newcommand{\PHclass}{\mathbf{PH}}
\newcommand{\PSPACE}{\mathbf{PSPACE}}
\newcommand{\RSPACE}{\mathbf{RSPACE}}
\newcommand{\DSPACE}{\mathbf{DSPACE}}
\newcommand{\imin}{{\mathit{min}}}
\newcommand{\imax}{{\mathit{max}}}
\newcommand{\uvec}{\vec{u}}
\newcommand{\vvec}{\vec{v}}
\newcommand{\xvec}{\vec{x}}
\newcommand{\Sigmap}{\Sigma^\prime}
\newcommand{\alphap}{\alpha^\prime}
\newcommand{\betap}{\beta^\prime}
\newcommand{\gammap}{\gamma^\prime}
\newcommand{\mup}{\mu^\prime}
\newcommand{\mupp}{\mu^{\prime\prime}}
\newcommand{\nubar}{{\overline{\nu}}}
\newcommand{\nubars}{\nubar^*}
\newcommand{\nus}{\nu^*}
\newcommand{\betas}{\beta^*}
\newcommand{\deltas}{\delta^*}
\newcommand{\deltap}{\delta^\prime}
\newcommand{\deltapp}{\delta^{\prime\prime}}
\newcommand{\lambdabar}{\overline{\lambda}}
\newcommand{\lambdap}{\lambda^\prime}
\newcommand{\lambdapp}{\lambda^{\prime\prime}}
\newcommand{\pip}{\pi^\prime}
\newcommand{\pipp}{{\pi^{\prime\prime}}}
\newcommand{\Psip}{\Psi^\prime}
\newcommand{\psip}{\psi^\prime}
\newcommand{\phip}{\phi^\prime}
\newcommand{\sigmap}{\sigma^\prime}
\newcommand{\etap}{\eta^\prime}
\newcommand{\etapp}{{\eta^{\prime\prime}}}
\newcommand{\epsilonp}{{\epsilon^\prime}}
\newcommand{\epsilonpp}{\epsilon^{\prime\prime}}
\newcommand{\epsilonppp}{\epsilon^{\prime\prime\prime}}
\newcommand{\ap}{a^\prime}
\newcommand{\bp}{b^\prime}
\newcommand{\cp}{c^\prime}
\newcommand{\up}{u^\prime}
\newcommand{\fp}{f^\prime}
\newcommand{\ep}{e^\prime}
\newcommand{\gp}{g^\prime}
\newcommand{\mpr}{m^\prime}
\newcommand{\np}{n^\prime}
\newcommand{\op}{o^\prime}
\newcommand{\qp}{q^\prime}
\newcommand{\rp}{r^\prime}
\newcommand{\gpp}{g^{\prime\prime}}
\newcommand{\epp}{e^{\prime\prime}}
\newcommand{\qpp}{q^{\prime\prime}}
\newcommand{\rpp}{r^{\prime\prime}}
\newcommand{\vp}{v^\prime}
\newcommand{\ypp}{y^{\prime\prime}}
\newcommand{\xpp}{x^{\prime\prime}}
\newcommand{\zp}{z^\prime}
\newcommand{\hp}{{h^\prime}}
\newcommand{\lp}{{l^\prime}}
\newcommand{\zpp}{{z^{\prime\prime}}}
\newcommand{\kp}{{k^\prime}}
\newcommand{\Dp}{{D^\prime}}
\newcommand{\Pp}{P^\prime}
\newcommand{\Ppp}{P^{\prime\prime}}
\newcommand{\Qp}{Q^\prime}
\newcommand{\Qpp}{Q^{\prime\prime}}
\newcommand{\Tp}{T^\prime}
\newcommand{\Ep}{E^\prime}
\newcommand{\Dbar}{\overline{D}}
\newcommand{\Lp}{L^\prime}
\newcommand{\Lpp}{L^{\prime\prime}}
\newcommand{\Lbar}{\overline{L}}
\newcommand{\Lhat}{\widehat{L}}
\newcommand{\Ltilde}{\tilde{L}}
\newcommand{\Lcap}{{L^\cap}}
\newcommand{\Lcup}{{L^\cup}}
\newcommand{\Lpbar}{\overline{\Lp}}
\newcommand{\Mp}{M^\prime}
\newcommand{\Mpp}{M^{\prime\prime}}
\newcommand{\Mbar}{\overline{M}}
\newcommand{\Np}{N^\prime}
\newcommand{\Npp}{N^{\prime\prime}}
\newcommand{\Rp}{R^\prime}
\newcommand{\xp}{x^\prime}
\newcommand{\yp}{y^\prime}
\newcommand{\Uhat}{\widehat{U}}
\newcommand{\Up}{U^\prime}
\newcommand{\Upp}{U^{\prime\prime}}
\newcommand{\Vp}{V^\prime}
\newcommand{\Vhat}{\widehat{V}}
\newcommand{\Vbar}{\overline{V}}
\newcommand{\Ap}{{A^\prime}}
\newcommand{\App}{{A^{\prime\prime}}}
\newcommand{\Cp}{C^\prime}
\newcommand{\Fp}{F^\prime}
\newcommand{\Gp}{G^\prime}
\newcommand{\Fpp}{F^{\prime\prime}}
\newcommand{\Zf}{{\Bbb{Z}}}
\newcommand{\Qf}{{\Bbb{Q}}}
\newcommand{\Rf}{{\Bbb{R}}}
\newcommand{\Cf}{{\Bbb{C}}}
\newcommand{\qacc}{q^{acc}}
\newcommand{\qrej}{q^{rej}}
\newcommand{\qnon}{q^{non}}
\newcommand{\Qadd}{Q_{add}}
\newcommand{\Qacc}{Q_{acc}}
\newcommand{\Qrej}{Q_{rej}}
\newcommand{\Qnon}{Q_{non}}
\newcommand{\Qhalt}{Q_{halt}}
\newcommand{\Qjunk}{Q_{junk}}
\newcommand{\Qaccp}{{Q_{acc}^\prime}}
\newcommand{\Qrejp}{{Q_{rej}^\prime}}
\newcommand{\Qnonp}{{Q_{non}^\prime}}
\newcommand{\Qhaltp}{{Q_{halt}^\prime}}
\newcommand{\Qjunkp}{{Q_{junk}^\prime}}
\newcommand{\Qaccpp}{{Q_{acc}^{\prime\prime}}}
\newcommand{\Qrejpp}{{Q_{rej}^{\prime\prime}}}
\newcommand{\Qnonpp}{{Q_{non}^{\prime\prime}}}
\newcommand{\Qjunkpp}{{Q_{junk}^{\prime\prime}}}
\newcommand{\Cacc}{C_{acc}}
\newcommand{\Crej}{C_{rej}}
\newcommand{\Cnon}{C_{non}}
\newcommand{\Eacc}{E_{acc}}
\newcommand{\Erej}{E_{rej}}
\newcommand{\Enon}{E_{non}}
\def\cent{{\hbox{\rm\rlap/c}}}
\newcommand{\centp}{{\cent}^\prime}
\newcommand{\Bra}[1]{{\langle{#1}|}}
\newcommand{\Ket}[1]{{|{#1}\rangle}}
\newcommand{\BraKet}[2]{{\langle{#1}|{#2}\rangle}}
\newtheorem{theorem}{{\bf Theorem}}[\mytheoremcounter]
\newtheorem{lemma}[theorem]{{\bf Lemma}}
\newtheorem{claim}[theorem]{{\bf Claim}}
\newtheorem{example}[theorem]{{\bf Example}}
\newtheorem{question}[theorem]{{\bf Question}}
\newtheorem{answer}[theorem]{{\bf Answer}}
\newtheorem{conjecture}[theorem]{{\bf Conjecture}}
\newtheorem{proposition}[theorem]{{\bf Proposition}}
\newtheorem{corollary}[theorem]{{\bf Corollary}}
\newtheorem{fact}[theorem]{{\bf Fact}}
\newtheorem{definition}[theorem]{{\bf Definition}}
\newtheorem{remark}[theorem]{{\bf Remark}}
\newtheorem{thoughts}[theorem]{{\bf Thoughts}}
\newenvironment{proof}{ \begin{trivlist} 
                        \item \vspace{-\topsep} \noindent{\bf Proof:}\ }
                      {\rule{5pt}{5pt}\end{trivlist}}
\newcommand{\Case}[2]{\noindent{\bf Case #1:}#2}
\newcommand{\Subcase}[2]{\noindent{\bf Subcase #1:}#2}
\newcommand{\Half}{\frac{1}{2}}
\newcommand{\RtHalf}{\frac{1}{\sqrt{2}}}
\newcommand{\cA}{{\mathcal{A}}}
\newcommand{\cC}{{\mathcal{C}}}
\newcommand{\cE}{{\mathcal{E}}}
\newcommand{\cF}{{\mathcal{F}}}
\newcommand{\cH}{{\mathcal{H}}}
\newcommand{\cI}{{\mathcal{I}}}
\newcommand{\cK}{{\mathcal{K}}}
\newcommand{\cL}{{\mathcal{L}}}
\newcommand{\cO}{{\mathcal{O}}}
\newcommand{\cP}{{\mathcal{P}}}
\newcommand{\cR}{{\mathcal{R}}}
\newcommand{\cS}{{\mathcal{S}}}
\newcommand{\cU}{{\mathcal{U}}}
\newcommand{\Span}{{\mathit{Span}}}
\newcommand{\Ch}[2]{{#1 \choose #2}}
\newcommand{\Ul}[1]{{\underline{#1}}}
\newcommand{\Floor}[1]{{\lfloor #1 \rfloor}}
\newcommand{\ignore}[1]{}
\newcommand{\noignore}[1]{#1}

\newcommand{\RMO}{\mathbf{RMO}}
\newcommand{\UMO}{\mathbf{UMO}}
\newcommand{\RMOe}{\mathbf{RMO}_\epsilon}
\newcommand{\RMM}{\mathbf{RMM}}
\newcommand{\UMM}{\mathbf{UMM}}
\newcommand{\RMMe}{\mathbf{RMM}_\epsilon}

\newcommand{\MOQFA}{\mathbf{MOQFA}}
\newcommand{\MOQFAe}{\mathbf{MOQFA}_\epsilon}
\newcommand{\MMQFA}{\mathbf{MMQFA}}
\newcommand{\MMQFAe}{\mathbf{MMQFA}_\epsilon}
\newcommand{\GQFA}{\mathbf{GQFA}}
\newcommand{\GQFAe}{\mathbf{GQFA}_\epsilon}

\newcommand{\REG}{\mathbf{REG}}
\newcommand{\PFA}{\mathbf{PFA}}
\newcommand{\PFAe}{\mathbf{PFA}_\epsilon}
\newcommand{\GFA}{\mathbf{GFA}}

\newcommand{\Foreach}[2]{\\{\bf\tt{for\ each}} $#1$ {\bf\tt{do}}\+ #2
\- \\ {\bf\tt{rof}}}
\newcommand{\Forloop}[2]{\\{\bf\tt{for}} $#1$ {\bf\tt{do}}\+ #2
\- \\ {\bf\tt{rof}}}
\newcommand{\Ifthen}[2]{\\{\bf\tt{if}} $#1$ {\bf\tt{then}}\+ #2
\- \\ {\bf\tt{fi}}}
\newcommand{\Ifelse}[3]{\\{\bf\tt{if}} $#1$ {\bf\tt{then}}\+ #2
\- \\ {\bf\tt{else}}\+ #3 \- \\ {\bf\tt{fi}}}
\newcommand{\Stmt}[1]{\\$#1$;}
\newcommand{\StartStmt}[1]{\+\kill$#1$;}
\newenvironment{pseudocode}{\begin{tabbing} 
\ \ \ \ \=\ \ \ \ \=\ \ \ \ \=\ \ \ \ \=\ \ \ \ \=\ \ \ \ \=\ \ \ \ \=\
\ \ \ \= } {\end{tabbing}}

\providecommand{\SaveProof}[3]{#3}
\providecommand{\SketchProof}[4]{#3}
\providecommand{\AppendixProof}[3]{}

\newcommand{\MoveProofsToAppendix}{\include{movemacs}}

\newcommand{\ShortSep}{\\ & &}
\newcommand{\LongSep}{}
\providecommand{\DefSep}{\LongSep}
\newcommand{\UseShortSep}{\renewcommand{\DefSep}{\ShortSep}}

\newcommand{\UseAbstract}[2]{#1}
\newcommand{\UseOtherAbstract}{\include{absselect}}

\newcommand{\StretchPage}{ \addtolength{\textheight}{0.05\textheight}
                           \addtolength{\topmargin}{-0.03\textheight}
                         }


\newcommand{\DoFigure}[4]{
                          \begin{figure}[ht]
                            \begin{center}
                              \ \psfig{file=#1,width=#2}\ 
                            \end{center}
                            \caption{#3\label{#4}}
                          \end{figure}
                         }


\bibliographystyle{alpha}


\title{Characterizations of 1-Way Quantum Finite Automata}
\author{Alex Brodsky \\ 
       	Department of Computer Science \\
        University of British Columbia \\ 
        {\tt abrodsky@cs.ubc.ca}
        \and 
        Nicholas Pippenger \\ 
       	Department of Computer Science \\
        University of British Columbia \\
        {\tt nicholas@cs.ubc.ca} }
\maketitle

\abstract{
\UseAbstract{
The 2-way quantum finite automaton introduced by Kondacs and
Watrous\cite{KoWa97} can accept non-regular languages with bounded
error in polynomial time.  If we restrict the head of the automaton to
moving classically and to moving only in one direction, the acceptance
power of this 1-way quantum finite automaton is reduced to a proper
subset of the regular languages.

In this paper we study two different models of 1-way quantum finite
automata.  The first model, termed measure-once quantum finite
automata, was introduced by Moore and Crutchfield\cite{MoCr98}, and the
second model, termed measure-many quantum finite automata, was
introduced by Kondacs and Watrous\cite{KoWa97}.

We characterize the measure-once model when it is restricted to
accepting with bounded error and show that, without that restriction, it
can solve the word problem over the free group.  We also show that it
can be simulated by a probabilistic finite automaton and describe an
algorithm that determines if two measure-once automata are equivalent.

We prove several closure properties of the classes of languages
accepted by measure-many automata, including inverse homomorphisms, and
provide a new necessary condition for a language to be accepted by the
measure-many model with bounded error.  Finally, we show that
piecewise testable sets can be accepted with bounded error by a
measure-many quantum finite automaton, in the process introducing new
construction techniques for quantum automata.
}{
In this paper we study two different models of 1-way quantum finite
automata.  The first model, termed measure-once quantum finite
automata, was introduced by Moore and Crutchfield \cite{MoCr98}, and the
second model, termed measure-many quantum finite automata, was
introduced by Kondacs and Watrous\cite{KoWa97}.
We characterize the measure-once model when it is restricted to
accepting with bounded error and show that, without that restriction, it
can solve the word problem over the free group.  We also show that it
can be simulated by a probabilistic finite automaton and describe an
algorithm that determines if two measure-once automata are equivalent.
We prove several closure properties of the classes of languages
accepted by measure-many automata, including inverse homomorphisms, and
provide a new necessary condition for a language to be accepted by the
measure-many model with bounded error.  Finally, we show that piecewise
testable sets can be accepted with bounded error by a measure-many
quantum finite automaton, in the process introducing new construction
techniques for quantum automata. 
} }

\section{Introduction}
In 1997 Kondacs and Watrous\cite{KoWa97} showed that a 2-way quantum
finite automaton (2QFA) could accept the language $L = a^nb^n$ in
linear time with bounded error.  The ability of the reading head to be
in a superposition of locations rather than in a single location at any
time during the computation gives the 2QFA its power.  Even if we
restrict the head of a 2-way quantum finite automaton from moving left,
we can still construct a 2QFA that can accept the language $\Lp =
\{x\in \{a,b\}^*\ |\ |x|_a = |x|_b\}$ in linear time with bounded
error.  However, if we restrict the head of a 2QFA to moving right on
each transition, we get the 1-way quantum finite automaton of Kondacs
and Watrous\cite{KoWa97}, which, when accepting with bounded error, can
only accept a proper subset of the regular languages.

If the reading head is classical then quantum mechanical evolution
hinders language acceptance; restricting the set of languages accepted
by 1-way quantum finite automata with bounded error to a proper subset
of the regular languages\cite{KoWa97}.

During its computation, a 1-way QFA performs measurements on its
configuration.  Since the acceptance capability of a 1-way QFA depends
on the measurements that the QFA may perform during the computation, we
investigate two models of 1-way QFAs that differ only in the type of
measurement that they perform during the computation.

The first model, termed measure-once quantum finite automata (MO-QFAs),
is similar to the one introduced by Moore and
Crutchfield\cite{MoCr98}.  The second model, termed measure-many
quantum finite automata (MM-QFAs), is similar to the one introduced by
Kondacs and Watrous\cite{KoWa97}, and is more complex than the MO-QFA.
The main difference between the two models is that a measure-once
automaton performs one measurement at the end of its computation, while
a measure-many automaton performs a measurement after every
transition.  This makes the measure-many model more powerful than the
measure-once model,  where the power of a model refers to the
acceptance capability of the corresponding automata.

First, we present results dealing with MO-QFAs.  We show that the class
of languages accepted by MO-QFAs with bounded error is exactly the
class of group languages.  Consequently, this class of
languages accepted by MO-QFAs is closed under inverse homomorphisms,
word quotients, and boolean operations.  We show that MO-QFAs that do
not accept with bounded error can accept non-regular languages and, in
particular, can solve the word problem over the free group.  We also
describe an algorithm that determines if two MO-QFAs are equivalent and
prove that probabilistic finite automata (PFAs) can simulate MO-QFAs.

Second, we shift our focus to MM-QFAs.  We show that the classes of
languages accepted by these automata are closed under complement,
inverse homomorphisms, and word quotients.  We prove by example that
the class of languages accepted by MM-QFAs with bounded error is not
closed under homomorphisms, and prove a necessary condition for
membership within this class.  We also relate the sufficiency of this
condition to the question of whether the class is closed under boolean
operations.  Finally, we show, by construction, that MM-QFAs can accept
piecewise testable sets with bounded error and introduce novel concepts 
for constructing MM-QFAs.

The rest of the paper is organized in the following way: Section~2
contains the definitions of the quantum automata and background
information, Section~3 discusses measure-once quantum finite automata,
Section~4 discusses measure-many quantum finite automata, and Section~5
summarizes.

\section{Definitions and Background}
\subsection{Definition of MO-QFA}
A measure-once quantum finite automaton is defined by a 5-tuple 
\[M = (Q,\Sigma,\delta,q_0,F)\]
where $Q$ is a finite set of states, $\Sigma$ is a finite input alphabet
with an end-marker symbol $\$$, $\delta$ is the transition function
 \[\delta: Q \times \Sigma \times Q \rightarrow \Cf\] 
that represents the probability density amplitude that flows from state
$q$ to state $\qp$ upon reading symbol~$\sigma$, the state~$q_0$ is
the initial configuration of the system, and $F$ is the set of accepting
states. For all states $q_1,q_2 \in Q$ and symbols $\sigma \in \Sigma$ 
the function $\delta$ must be unitary, thus satisfying the condition
\begin{equation}\label{restriction}
\sum_{\qp \in Q}\overline{\delta(q_1,\sigma,\qp)}\delta(q_2,\sigma,\qp)
 = \left\{\begin{array}{lr}
         1 & q_1 = q_2 \\
         0 & q_1 \not= q_2\\
         \end{array}\right..
\end{equation}
We assume that all input is terminated by the end-marker $\$$; this is
the last symbol read before the computation terminates.  At the end of
a computation $M$ measures its configuration; if it is in an accepting
state then it accepts, otherwise it rejects.  This definition is
equivalent to that of the QFA defined by Moore and
Crutchfield\cite{MoCr98}.

The configuration of $M$ is a linear superposition of states and is
represented by an $n$-dimensional complex unit vector, where $n =
|Q|$.  This vector is denoted by
 \[\Ket{\Psi} = \sum^n_{i=1}\alpha_i\Ket{q_i}\] 
where $\{\Ket{q_i}\}$ is the set orthonormal basis vectors
corresponding to the states of $M$.  The coefficient $\alpha_i$ is the
probability density amplitude of $M$ being in state~$q_i$.  Since
$\Ket{\Psi}$ is a unit vector, it follows that
$\sum^n_{i=1}|\alpha_i|^2 = 1$.

The transition function $\delta$ is represented by a set of unitary
matrices $\{U_\sigma\}_{\sigma\in \Sigma}$ where $U_\sigma$ represents
the unitary transitions of $M$ upon reading symbol~$\sigma$.  If $M$ is
in configuration $\Ket{\Psi}$ and reads symbol~$\sigma$ then the new
configuration of $M$ is denoted by
 \[\Ket{\Psip} = U_\sigma\Ket{\Psi} = \sum_{q_i,q_j\in Q}
		       \alpha_i\delta(q_i,\sigma,q_j)\Ket{q_j}.\]
Measurement is represented by a diagonal zero-one projection matrix
$P$ where $P_{ii} = [q_i \in F]$.  The probability of $M$ accepting 
string~$x$ is defined by
 \[p_M(x) = \Bra{\Psi_x}P\Ket{\Psi_x} = \|P\Ket{\Psi_x}\|^2 \] 
where
$\Ket{\Psi_x} = U(x)\Ket{q_0} = U_{x_n}U_{x_{n-1}}...U_{x_1}\Ket{q_0}$.

\subsection{Definition of MM-QFA}
A measure-many quantum finite automaton is defined by a 6-tuple
\[M=(Q,\Sigma,\delta,q_0,\Qacc,\Qrej)\]
where $Q$ is a finite set of states, $\Sigma$ is a finite input
alphabet with an end-marker symbol~$\$$, $\delta$ is a unitary
transition function of the same form as for an MO-QFA, and the
state~$q_0$ is the initial configuration of $M$.  The set $Q$ is
partitioned into three subsets:  $\Qacc$ is the set of
halting accepting states, $\Qrej$ is the set of halting
rejecting states, and $\Qnon$ is the set of non-halting states.

The operation of an MM-QFA is similar to that of an MO-QFA except that
after every transition $M$ measures its configuration with respect to
the three subspaces that correspond to the three subsets $\Qnon$,
$\Qacc$, and $\Qrej$: $\Enon = \Span(\{\Ket{q}\ |\  q \in \Qnon\})$,
$\Eacc = \Span(\{\Ket{q}\ |\  q \in \Qacc\})$, and $\Erej =
\Span(\{\Ket{q}\ |\  q \in \Qrej\})$.  If the configuration of $M$ is
in $\Enon$ then the computation continues; if the configuration is in
$\Eacc$ then $M$ accepts, otherwise it rejects.  After every
measurement the superposition collapses into the measured subspace and
is renormalized.

Just like MO-QFAs, the configuration of an MM-QFA is represented by a
complex $n$-dimensional vector, the transition function is represented
by unitary matrices, and measurement is represented by diagonal
zero-one projection matrices that project the vector onto the
respective subspaces.

The definition of an MM-QFA is almost identical to the definition by
Kondacs and Watrous in\cite{KoWa97}.  The only difference is that we
only require one end-marker at the end of the tape, rather than two
end-markers, at the start and end of the tape; this does not affect the
acceptance power of the automaton; see Appendix~\ref{equiv_mmqfa} for
further details.

Since $M$ can have a non-zero probability of halting part-way through
the computation, it is useful to keep track of the cumulative accepting
and rejecting probabilities.  Therefore, in some cases we use the
representation, of Kondacs and Watrous\cite{KoWa97} that represents the
state of $M$ as a triple $(\Ket{\Psi},p_{acc},p_{rej})$,
where $p_{acc}$ and $p_{rej}$ are the cumulative probabilities of
accepting and rejecting.  The evolution of $M$ on reading
symbol~$\sigma$ is denoted by
 \[(P_{non}\Ket{\Psip},\ p_{acc} + \|P_{acc}\Ket{\Psip}\|^2,\ 
			       p_{rej} + \|P_{rej}\Ket{\Psip}\|^2)\]
where $\Ket{\Psip} = U_\sigma\Ket{\Psi}$, and $P_{acc}$, $P_{rej}$, and
$P_{non}$ are the diagonal zero-one projection matrices that project the
configuration onto the non-halting, accepting and rejecting subspaces.

\subsection{Language Acceptance}
A QFA $M$ is said to accept a language $L$ with cut-point $\lambda$ if
for all $x \in L$ the probability of $M$ accepting $x$ is greater than
$\lambda$ and for all $x \not\in L$ the probability of $M$ accepting
$x$ is at most $\lambda$.  A QFA $M$ accepts $L$ with bounded error if
there exists an $\epsilon > 0$ such that for all $x \in L$ the
probability of $M$ accepting $x$ is greater than $\lambda + \epsilon$
and for all $x \not\in L$ the probability of $M$ accepting $x$ is less
than $\lambda - \epsilon$.  We call $\epsilon$ the margin.  

We partition the languages accepted by QFAs into several natural
classes.  Let the class $\RMOe$ be the set of languages accepted by an
MO-QFA with margin of at least $\epsilon$.  Let the restricted class of
languages, $\RMO = \cup_{\epsilon > 0}\RMOe$, be the set of languages
accepted by an MO-QFA with bounded error, and let the unrestricted
class of languages, $\UMO = \RMO_0$, be the set of languages accepted
by an MO-QFA with unbounded error.  We define the languages classes 
$\RMMe$, $\RMM$ and $\UMM$ accepted by an MM-QFA in a similar fashion.

Since the cut-point of a QFA can be arbitrarily raised or lowered, we 
could without loss of generality fix the cut-point to be $\Half$.
However, for the purposes of presentation we use the general cut-point
definition stated above.

\subsection{Reversible Finite Automata}
Unitary operations are reversible, thus QFAs bear strong resemblance to
various variants of reversible finite automata.  A group finite
automaton (GFA) is a deterministic finite automata (DFA) $M =
(Q,\Sigma,\delta,q_0,F)$ with the restriction that for every state~$q
\in Q$ and every input symbol $\sigma \in \Sigma$ there exists exactly
one state~$\qp \in Q$ such that $\delta(\qp,\sigma) = q$, i.e. $\delta$
is a complete one-to-one function and the automaton derived from $M$ by
reversing all transitions is deterministic.

A reversible finite automata (RFA) is a DFA $M =
(Q,\Sigma,\delta,q_0,F)$ such that for every state~$q \in Q$ and for
every symbol $\sigma \in \Sigma$ there is at most one state~$\qp \in Q$
such that $\delta(\qp,\sigma) = q$, or, if there exist distinct states
$q_1,q_2 \in Q$ and symbol $\sigma \in \Sigma$ such that
$\delta(q_1,\sigma) = q = \delta(q_2,\sigma)$, then $\delta(q,\Sigma) =
\{q\}$.  The latter type of state is called a spin state because once
an RFA enters it, it will never leave it.  This definition is
equivalent to the one used by Ambainis and Freivalds\cite{AmFr98} and
is an extension of Pin's\cite{Pi87} definition.

\subsection{Previous Work}
Moore and Crutchfield\cite{MoCr98} introduced a variant of the MO-QFA
model and investigated the model in terms of quantum regular languages
(QRLs).  They showed several closure properties including closure under
inverse homomorphisms and derived a method for bilinearizing the
representation of an MO-QFA that transforms it into a generalized 
stochastic system.

Kondacs and Watrous\cite{KoWa97} introduced a variant of the MM-QFA
that was derived from their 2QFA model.  Using a technique similar to
Rabin's\cite{Ra63}, Kondacs and Watrous proved that 1-way QFAs that
accept with bounded error are restricted to accepting a proper subset
of the regular languages and that the language $L = \{a,b\}^*b$ is not
a member of that subset.

Ambainis and Freivalds\cite{AmFr98} showed that MM-QFAs could accept
languages with probability higher than $\frac{7}{9}$ if and only if the
language could be accepted by an RFA, which is equivalent to being
accepted with certainty by an MM-QFA.  In~\cite{AmBoFrKi99} Ambainis,
Bonner, Freivalds, and Kikusts, construct a hierarchy of languages such
that the $i$th language in the hierarchy can be accepted by a MM-QFA
with at most probability $p_i$, where the series $(p_i)$ converges to
$\Half$ and is strictly decreasing.

Ambainis, Nayak, Ta-Shma, and Vazirani\cite{AmNaTaVa99}, and 
Nayak\cite{Na99}, investigated how efficiently MM-QFAs can be constructed
compared to DFAs.  They showed that for some languages the 
accepting MM-QFA is exponentially larger than the corresponding DFA.

In \cite{AmIw99} Amano and Iwama studied a restricted version of the 2QFA
model where the head was not allowed to move right.  They showed that 
the emptiness problem for this model is undecidable.   This is another
instance where quantum mechanics provides computational power that is not
achievable in the classical case.

\section{MO-QFAs}
\subsection{Bounded Error Acceptance}
\label{sec_char_moqfa}
The restriction that MO-QFAs accept with bounded error is as limiting
as in the case of PFAs\cite{Ra63}.  Since MM-QFAs can only accept a
proper subset of the regular languages if they are required to accept
with bounded error and since every MO-QFA can be simulated exactly by
an MM-QFA, the class $\RMO$ is a proper subset of the regular
languages.  The class $\RMO$ is exactly the class of languages accepted
by group finite automata (GFAs), otherwise known as group languages,
and whose syntactic semigroups are groups, see Eilenberg\cite{Ei76}.
This result is implied by Theorem 7 in~\cite{MoCr98} but is not stated
in the paper.  To prove this result we first need Lemma~\ref{u_id}.

\begin{lemma}\label{u_id} 
Let $U$ be a unitary matrix.  For any $\epsilon > 0$ there exists an
integer $n > 0$ such that for all vectors $x$, where $\|x\|^2 \leq 1$,
it is true that $\|(I - U^n)x\|^2 < \epsilon$.
\end{lemma}
\SaveProof{\lemmauidproof}{lemma_u_id}{\begin{proof}
Let $m = dim(U)$.  Since $U$ is a normal matrix, $U^n$ can be written 
as
\[U^n = P D^n P^{-1}\]
where $P$ is a unitary matrix and $D$ is the diagonal matrix of
eigenvalues with the $j$th eigenvalue having the form $e^{i\pi
r_j}$\cite{Or87}.  If all eigenvalues in $D$ are rotations through
rational fractions of $\pi$, i.e.  $r_j$ is rational, then let $n =
2\prod_{j=1}^m q_j$ where $q_j$ is the denominator of $r_j$.  Thus $D^n
= I$ and we are done.

Otherwise, at least one eigenvalue is a rotation of unity through an
irrational fraction of $\pi$.  Let $l \leq m$ be the number of these
eigenvalues.  For the other $m-l$ eigenvalues compute $n$, just as
above, and let $\Dp = D^n$.  The value of the $j$th element on the
diagonal of $\Dp$ is either $1$ or $e^{i\pi nr_j}$ where $r_j$ is some
irrational real number.  Consider taking $\Dp$ to some power $k \in
\Zf^+$.  The values that are $1$ do not change, but the other $l$
values that are of the form $e^{i\theta_jk}$ where $\theta_j = \pi n
r_j$, form a vector that varies through a dense subset in an
$l$-dimensional torus.  Hence, there exists $k$ such that the
$l$-dimensional vector is arbitrarily close to $\vec{1}$.  Thus, for 
any $\epsilonp > 0$ there exists a $k > 0$ such that $\|(I -
\Dp^k)\vec{1}\|^2 < \epsilonp$.  Hence
\begin{eqnarray*}
\|(I - U^{nk})x\|^2 & =    & \|(I - PD^{nk}P^{-1})x\|^2 \\
                    & =    & \|P(I - \Dp^k)P^{-1}x\|^2 \\
                    & \leq & \|(I - \Dp^k)m\vec{1}\|^2 \\
                    & =    & m^2\|(I - \Dp^k)\vec{1}\|^2\\
                    & \leq & m^2\epsilonp.
\end{eqnarray*}
Select $\epsilonp$ such that $\epsilonp < \frac{\epsilon}{m^2}$
to complete the proof.
\end{proof}}

Lemma~\ref{measure}, due to Bernstein and Vazirani\cite{BeVa97}, states
that if two configurations are close, then the differences in
probability distributions of the configurations is small.  This lemma
relates the closeness of configurations to the variation distance
between their probability distributions and allows us to partition the
set of reachable configurations into equivalence classes.  The variation
distance between two probability distributions is the maximum difference
in the probabilities of the same event occurring with respect to both
distributions.

\begin{lemma}[Bernstein and Vazirani, 1997]\label{measure}\ \\
Let $\Ket{\psi}$ and $\Ket{\varphi}$ be two complex vector such that
$\|\,\Ket{\psi}\|^2 = \|\,\Ket{\varphi}\|^2 = 1$ and $\|\,\Ket{\psi} -
\Ket{\varphi}\|^2 < \epsilon$.  The total variation distance
between the probability distributions resulting from measurement of
$\Ket{\psi}$ and $\Ket{\varphi}$ is at most $4\epsilon$.
\end{lemma}

Theorem~\ref{moqfagfa} follows from these two lemmas.

\begin{theorem}\label{moqfagfa}
A language $L$ can be accepted by an MO-QFA with bounded error if and
only if it can be accepted by a GFA.
\end{theorem}
\SaveProof{\theoremmoqfagfaproof}{theorem_moqfagfa}{\begin{proof}
The `if' direction follows from the fact that the transition function
for a GFA is also a valid transition function for an MO-QFA that can
accept the same language with certainty.

For the `only if' direction, by contradiction, assume that there exists
a language $L$ that can be accepted by an MO-QFA with bounded error but
cannot be accepted by a GFA.  Since the class $\RMO$ is a subset of the
regular languages, $L$ must be regular.  Let $M =
(Q,\Sigma,\delta,q_0,F)$ be an MO-QFA that accepts $L$ with bounded
error.  If two strings $x$ and $y$ take $M$ into the same reachable
configuration, then for any string $z$ the probability of $M$ accepting
$xz$ is equal to the probability of $M$ accepting $yz$, which means
that $xz \in L$ if and only if $yz \in L$.  Therefore, the space of
reachable configurations of $M$'s computation can be partitioned into a
finite number of equivalence classes defined by the corresponding
minimal DFA for $L$.

Let $\Ket{\psi}$ and $\Ket{\varphi}$ denote reachable configurations of
$M$ and let $\sim_L$ be the right invariant equivalence relation
induced by $L$.  Since $L$ cannot be accepted by a GFA, there must
exist two distinct equivalence classes $[y]$ and $[\yp]$, an
equivalence class $[x]$, and a symbol $\sigma \in \Sigma$, such that
$[y\sigma] \sim_L [\yp\sigma] \sim_L [x]$.  If $U_\sigma$ is the
transition matrix for symbol $\sigma$, $\Ket{\psi} \in [y]$ and
$\Ket{\varphi} \in [\yp]$ then $U_\sigma\Ket{\psi} \in [x]$ and
$U_\sigma\Ket{\varphi} \in [x]$.

Since $M$ accepts $L$ with bounded error, let $\epsilon$ be the
margin.  By Lemma~\ref{u_id} there exists an integer $k > 0$ such that
$\|(I - U^k_\sigma)\Ket{\psi}\|^2 < \frac{\epsilon}{4}$ and $\|(I -
U^k_\sigma)\Ket{\varphi}\|^2 < \frac{\epsilon}{4}$.  Hence,
$U^k_\sigma\Ket{\psi} \in [y]$ because if
\begin{eqnarray*}
\|(I-U^k_\sigma)\Ket{\psi}\|^2 &=& \|\,\Ket{\psi}-U^k_\sigma\Ket{\psi}\|^2 \\
                               &=& \|V(\Ket{\psi}-U^k_\sigma\Ket{\psi})\|^2\\
                               &<& \frac{\epsilon}{4}
\end{eqnarray*}
where $V$ is an arbitrary unitary matrix, then by Lemma~\ref{measure}
the probability of $VU^k_\sigma\Ket{\psi}$ being measured in a
particular state is within $\epsilon$ of $V\Ket{\psi}$ being measured
in the same state; this probability is less than the margin.  Similarly
$U^k_\sigma\Ket{\varphi} \in [\yp]$.  Hence $[y] \sim_L [y\sigma^k]$
and $[\yp] \sim_L [\yp\sigma^k]$.

We assumed that $[x] \sim_L [y\sigma] \sim_L [\yp\sigma]$ and showed that
$[y] \sim_L [y\sigma^k]$ and $[\yp] \sim_L [\yp\sigma^k]$; therefore,
$[y] \sim_L [x\sigma^{k-1}] \sim_L [\yp]$.  Let $z$ be the string that
distinguishes $[y]$ and $[\yp]$.  Then the string $\sigma^{k-1}z$
partitions $[x]$ into at least two distinct equivalence classes, but
this is a contradiction.  Therefore, there cannot exist a language $L$ that
can be accepted by an MO-QFA with bounded error but not by a GFA.
\end{proof}}

Theorem~\ref{moqfagfa} implies that $\RMO_\epsilon = \RMO_\epsilonp$
for all $\epsilon, \epsilonp > 0$, hence there are most two distinct
classes of languages accepted by MO-QFAs, the restricted class $\RMO$,
which is equivalent to the class of languages accepted by a GFA, and
the unrestricted class $\UMO$.

It follows immediately from Theorem~\ref{moqfagfa} that the class
$\RMO$ is closed under boolean operations, inverse homomorphisms, and
word quotients, and is not closed under homomorphisms.  

\subsection{Non-Regular Languages}
Unlike the class $\RMO$, the class $\UMO$ contains languages that are
non-regular.  This is not surprising given that Rabin\cite{Ra63} proved
a similar result for PFAs.  In fact our proof closely mimics
Rabin's\cite{Ra63} technique.

\begin{lemma}\label{lem_non_reg}
Let $L = \{x \in \{a,b\}^*\ \vert\ |x|_a \not= |x|_b\}$, there exists a 
2-state MO-QFA $M$ that accepts $L$ with cut-point 0.
\end{lemma}
\begin{proof}
Let $M = (Q,\Sigma,\delta,q_0,F)$ where $Q = \{q_0,q_1\}$, $\Sigma =
\{a,b\}$, $F = \{q_1\}$, and $\delta$ is defined by the transition
matrices
\[U_a = U^{-1}_b = \left[\begin{array}{cc}
                         \cos\alpha & \sin\alpha \\
                        -\sin\alpha & \cos\alpha \\
                         \end{array}\right]\]
where $\alpha$ is an irrational fraction of $\pi$.  Since $U_a$ is a
rotation matrix and $\alpha$ is an irrational fraction of $\pi$, the
orbit formed by applying $U_a$ to $\Ket{q_0}$ is dense in the circle, 
and there exists only one $k$, such that $U^k_a\Ket{q_0} = \Ket{q_0}$,
namely $k = 0$.  This also holds for $U_b = U_a^{-1}$.  Thus,
$U(x)\Ket{q_0} = \Ket{q_0}$ if and only if the number of $U_a$
rotations applied to $\Ket{q_0}$ is equal to the number of $U_b$
rotations, which is true if and only if the $|x|_a = |x|_b$.
Otherwise, $M$ has a non-zero probability of halting in state $q_1$.
\end{proof}

Lemma~\ref{lem_non_reg} implies that the class $\RMO$ is properly
contained within the class $\UMO$ and therefore the two classes are
distinct.

The MO-QFA in Lemma~\ref{lem_non_reg} solves the word problem for the
infinite cyclic group: is the input word equal to the identity element
in the group, where the group has only one generator element, say $a$,
and its inverse $b = a^{-1}$.  We can generalize this result to the
general word problem for the free group.  The word problem for a free
group is to decide whether or not a product of a sequence of elements
of the free group reduces to the identity\cite{LiZa77}.

\begin{lemma}\label{fg_qfa}
The word problem for the free group language can be accepted by an
MO-QFA.
\end{lemma}
\begin{proof}
Construct a free group of rotation matrices drawn from the group
$\mathrm{SO}_3$ as discussed by Wagon\cite{Wa85}.  Let $M =
(Q,\Sigma,\delta,q_0,F)$ be a $3$-state MO-QFA  where $\Sigma =
\{a,a^{-1},b,b^{-1},...\}$ such that $|\Sigma|$ is equal to the sum of
the number of rotation matrices and their inverses, $\delta$ is defined
by the rotation matrices and their inverses, and $F = \{q_0\}$.  The
MO-QFA will accept identity words with certainty and reject
non-identity words with a strictly non-zero probability, hence solving
the word problem for the free group.
\end{proof}

\subsection{Equivalence of MO-QFAs}
In classical automata theory there is an algorithm to determine if two
automata are equivalent.  We say that QFAs $M$ and $\Mp$ are
equivalent if their probability distributions over $\Sigma^*$ are the
same: for every word $x \in \Sigma$, the probability of $M$
accepting $x$ is equal to the probability of $\Mp$ accepting $x$.  In
order to determine if two MO-QFAs are equivalent we first bilinearize
them using the method detailed by Moore and Crutchfield\cite{MoCr98};
this yields two generalized stochastic systems.  We then apply
Paz's\cite[Page 21, Page 140]{Pa71} method for testing stochastic system 
equivalence to the generalized stochastic systems to determine if they 
have the same distribution.

\subsection{Simulation of MO-QFAs by PFAs}
Most classical computation is either deterministic or probabilistic,
hence it is useful to ask how probabilistic automata compare to
their quantum analogs.  In the case of MO-QFAs, any language accepted
by an MO-QFA can also be accepted by a PFA.  If $L$ can be accepted by
an MO-QFA with bounded error, then it can also be accepted by a PFA with
bounded error.

\begin{theorem}\label{pfa_sim_qfa}
Let $M$ be an MO-QFA that accepts $L$ with cut-point $\lambda$ then:
\begin{enumerate}
\item	There exists a PFA that accepts $L$ with some cut-point $\lambdap$.
\item	If $M$ accepts $L$ with bounded error, then there exists a PFA
        that accepts $L$ with bounded error.
\end{enumerate}
\end{theorem}
\begin{proof}
The second result follows from Theorem~\ref{moqfagfa} because every GFA is
also a PFA.

Since we can bilinearize $M$, $L$ is a generalized cut-point event
(GCE)\cite[Page 153]{Pa71}.  Since the class of GCEs is equal to the
class of probabilistic cut-point events (PCEs)\cite[Page 153]{Pa71},
which are accepted by PFAs,  there exists a PFA that can accept $L$
with some cut-point $\lambdap$.  
\end{proof}

Combining Theorem~\ref{pfa_sim_qfa} with Lemma~\ref{fg_qfa} yields a new
insight into the languages accepted by PFAs:

\begin{corollary}
The word problem for the free group language can be solved by a PFA.
\end{corollary}

\section{MM-QFAs}
Measure-many quantum finite automata are more powerful than MO-QFAs
because a measurement is performed after every transition.  This allows
the machine to terminate before reading the entire string and simulate
the spin states of RFAs.  

As mentioned before, an MM-QFA uses one end-marker while the Kondacs
and Watrous \cite{KoWa97} 1-way QFA  uses two end-markers.  The second
marker does not add any more power to the model, see
Appendix~\ref{equiv_mmqfa}, but makes constructing an MM-QFA easier
because the MM-QFA can start in an arbitrary configuration.  Hence, for
the sake of conciseness and clarity we shall assume that some of the
MM-QFAs constructed in the following proofs have two end-markers.

\subsection{Closure Properties}
Unlike the closure properties of the classes $\RMO$ and $\UMO$, which
can be derived easily, the closure properties of the classes $\RMM$ and
$\UMM$ are not as evident and in one important case unknown.  We show
that the classes $\RMM$ and $\UMM$ are  closed under complement,
inverse homomorphism and word quotient.  Similar to the
class $\RMO$, the class $\RMM$ is not closed under homomorphisms.  It
remains an open problem to determine whether the classes $\RMM$ and
$\UMM$ are closed under boolean operations.

Theorem~\ref{mm_close} proves that both classes are closed under
complement and inverse homomorphisms by showing that each class $\RMMe$
is closed under complement and inverse homomorphisms; closure under
word quotient follows directly from the latter, given the presence of 
end-markers.

\begin{theorem}\label{mm_close}
The class $\RMMe$ is closed under complement, inverse homomorphisms, and
word quotient.
\end{theorem}
\SaveProof{\theoremmmcloseproof}{theorem_mm_close}{\begin{proof}
Closure under complement follows from the fact that we can exchange the
accept and reject states of the MM-QFA.  This exchanges the
probabilities of acceptance and rejection but does not affect the
margin.

Given an MM-QFA $M$ and a homomorphism $h$ we construct an MM-QFA $\Mp$ 
that accepts $h^{-1}(L)$.  Let $M = (Q,\Sigma ,\delta,\Qacc,\Qrej)$ and
$\Mp = (\Qp,\Sigma,\deltap,\Qaccp,\Qrejp)$.  Assume that $\delta$
and $\deltap$ are defined in terms of matrices $\{U_\sigma\}_{\sigma \in
\Sigma}$ and $\{\Up_\sigma\}_{\sigma \in \Sigma}$.  Unlike the proof for
MO-QFAs in~\cite{MoCr98}, the direct construction of
\[\Up_\sigma = U(h(\sigma))\]
will not work because a measurement occurs between transitions, and
combining transitions without taking this into account could produce
incorrect configurations.  After every transition some amount of
probability amplitude is placed in the halting states and should not be
allowed to interact with the non-halting states in the following
transitions.  This is achieved by storing the amplitude in additional
states; this technique is also used in~\cite{AmNaTaVa99}.  Assume 
without loss of generality that
\begin{eqnarray*}
\Qnon  & = & \{q_i \in Q\ |\ 0 \leq i < n_{non} \} \\
\Qhalt & = & \{q_i \in Q\ |\ n_{non} \leq i < n \}
\end{eqnarray*}
where $n = |Q|$ and $n_{non} = |Q_{non}|$.  Let
$m = \max_{\sigma \in \Sigma}\{|h(\sigma)|\}$ and let
\[\Qp = Q \cup \Qhaltp\]
where
\begin{eqnarray*}
\Qhaltp &=& \{q_i\}^{n + m(n - n_{non})}_{i = n + 1}\\
\Qaccp  &=& \Qacc\cup\{q_{n+j(i-n_{non})} \in \Qhaltp\ |\ \DefSep
            q_i \in \Qacc, 1 \leq j \leq m\} \\
\Qrejp  &=& \Qrej\cup\{q_{n+j(i-n_{non})} \in \Qhaltp\ |\ \DefSep
            q_i \in \Qrej, 1 \leq j \leq m\}.
\end{eqnarray*}
Intuitively, we replicate the halting states $m$ times; each replication
is termed a halting state set.

We construct $\deltap$ from the matrices of $\delta$.  Let
$V_\sigma$ be a unitary block matrix
\[V_\sigma = U_{shift} \left[\begin{array}{cc}
                               U_\sigma &     \\
                                        & I_{m(n - n_{non})} \\
                               \end{array}\right]\]
where
\[U_{shift} = \left[ \begin{array}{ccc}
                I_{n_{non}} &                     & \\
                            &                     & I_{n - n_{non}} \\
                            & I_{m(n - n_{non})} & \\
              \end{array} \right].\]
The matrix $U_{shift}$ is a unitary matrix that shifts the amplitudes
in the halting set $i$ to the halting set $i+1$ and the amplitude in
halting set $m$ to halting set $0$.  
In analogy to the MO-QFA case where $\Up_\sigma =
U(h(\sigma))$, for MM-QFAs let
\[\Up_\sigma = V(h(\sigma)) = V_{x_k}V_{x_{k-1}}...V_{x_1}\]
where $h(\sigma) = x = x_1x_2...x_k$ and $k \leq m$.

After every $x_i$ sub-transition the halting amplitude is shifted and
stored in the $m+1$ halting sets of states.  When the sub-transition is
done, the amplitude in halt state set $0$ is zero, which is what is
required to prevent unwanted interactions.  A minimum of $m$
sub-transitions must occur before halting set $m$ contains non-zero
amplitude, but no more than $m$ sub-transitions will ever occur;
therefore halting set $0$ will never receive non-zero amplitude from
halting set $m$.  Since $\Mp$ has the same distribution as $M$, the
margin will not decrease.

Closure under word quotient follows from closure under inverse
homomorphism and the presence of both end-markers.
\end{proof}}

Just like the class $\RMO$, the class $\RMM$ is not closed
under homomorphisms.

\begin{theorem}\label{mm_homo}
The class $\RMM$ is not closed under homomorphisms.
\end{theorem}
\begin{proof}
Let $L = \{a,b\}^*c$ and define a homomorphism $h$ to be
$h(a) = a$, $h(b) = b$, and $h(c) = b$.  Since $L$ can 
be accepted by an RFA, $L \in \RMM$~\cite{AmFr98}, but $h(L) =
\{a,b\}^*b \not\in \RMM$, the result follows.
\end{proof}

A more interesting question is whether the classes $\RMM$ and $\UMM$
are closed under boolean operations.  Unlike MO-QFAs that have two
types of states: accept and reject, MM-QFAs have three types of states:
accept, reject, and non-halt.  Consequently, the standard procedure
of taking the tensor product of two automata to obtain their
intersection or union does not work.  A general method of intersecting
two MM-QFAs is not known.  Thus, it is not known whether $\RMM$ and
$\UMM$ are closed under boolean operations.

\subsection{Bounded Error Acceptance}
The restriction of bounded error acceptance reduces the class of
languages that an MM-QFA can accept to a proper subclass of the regular
languages\cite{KoWa97}.  To study the languages in class $\RMM$, we look
at their corresponding minimal automata.  Ambainis and
Freivalds\cite{AmFr98} showed that if the minimal DFA $M(L) =
(Q,\Sigma,\delta,q_0,F)$ contains an irreversible construction, defined
by two distinct states $q_1,q_2 \in Q$ and strings $x,y,z \in \Sigma^*$
such that $\delta(q_1,x) = \delta(q_2,x) = q_2$, $\delta(q_2,y) \in F$
and $\delta(q_2,z) \not\in F$, then an RFA cannot accept $L$ and an
MM-QFA cannot accept it with a probability greater than $\frac{7}{9}$;
this condition is both sufficient and necessary.

We derive a similar necessary condition for a language $L$ to be a
member of the class $\RMM$.  This condition, called the partial order
condition, is a relaxed version of a condition defined by Meyer and
Thompson\cite{MeTh69}.  A language $L$ is said to satisfy the partial
order condition if the minimal DFA for $L$ satisfies the partial order
condition.  A DFA satisfies the partial order condition if it does not
contain two distinguishable states $q_1,q_2 \in Q$ such that there
exists strings $x,y \in \Sigma^+$ where $\delta(q_1,x) =
\delta(q_2,x) = q_2$, and $\delta(q_2,y) = q_1$.  States $q_1$ and
$q_2$ are said to be distinguishable if there exists a string $z \in
\Sigma^*$ such that $\delta(q_1,z) \in F$ and $\delta(q_2,z) \not\in F$
or vice versa\cite{HoUl79}.  Using a result in~\cite{KoWa97},
Theorem~\ref{po_cond} proves that the partial order condition is
necessary for an MM-QFA to accept $L$ with bounded error.

\begin{theorem}\label{po_cond}
If $M = (Q,\Sigma,\delta,q_0,F)$ is a minimal DFA for language $L$
that does not satisfy the partial order condition then $L \not\in
\RMM$.
\end{theorem}
\SaveProof{\theorempocondproof}{theorem_po_cond}{ \begin{proof}
By contradiction, assume that $L \in \RMM$.  Let $L_b = \{a,b\}^*b$.
Since the minimal DFA for $L$ does not satisfy the partial order
condition there exist states $q_1,q_2 \in Q$ and strings $x,y \in
\Sigma^+$ as defined above and a distinguishing string $z \in \Sigma^*$
such that $\delta(q_1,z) \not\in F$ if and only if $\delta(q_2,z) \in
F$.  Without loss of generality assume that $\delta(q_1,z) \not\in F$
and $\delta(q_2,z) \in F$.

Let $s$ be the shortest string such that $\delta(q_0,s) = q_1$.  Let
$\Lp = s^{-1}Lz^{-1}$.  By Theorem~\ref{mm_close}, $\Lp \in \RMM$.
Define the homomorphism $h$ as
\begin{eqnarray*}
h(a)                & = & xy \\
h(b)                & = & x \\
h(\Sigma - \{a,b\}) & = & xy,
\end{eqnarray*}
where the last definition is for completeness.  Let $\Lpp =
h^{-1}(\Lp)$.  By Theorem~\ref{mm_close} $\Lpp \in \RMM$.
But $\Lpp = L_b \not\in \RMM$, a contradiction.
\end{proof}}

The partial order condition is so named because once the state $q_2$ is
visited, there is no path back to state $q_1$.  Thus, there exists a
partial order on the states of the DFA.  We do not know whether this
condition is also sufficient for MM-QFA acceptance with bounded error.
While we do not know whether the class $\RMM$ is closed under boolean
operations, Theorem~\ref{po_int} relates closure under intersection to
the partial order condition.

\begin{lemma}\label{po_min} 
Let $M$ be a DFA that satisfies the partial order condition.  The
minimal DFA $\Mp$ that accepts $L(M)$ satisfies the partial order
condition.
\end{lemma}
\SaveProof{\lemmapominproof}{lemma_po_min}{\begin{proof}
Let $M = (Q,\Sigma,\delta,q_0,F)$ be a DFA and $\Mp =
(\Qp,\Sigma,\deltap,\qp_0,\Fp)$ be the corresponding minimal DFA.
Assume by contradiction that $\Mp$ does not satisfy the partial order
condition.  Hence, $\Mp$ has two states that correspond to the
equivalence classes $[\qp_1]$ and $[\qp_2]$ such that $[\qp_1]x \sim_L
[\qp_2]x \sim_L [\qp_2]$ and $[\qp_2]y \sim_L [\qp_1]$.  By the
Myhill-Nerode theorem\cite{HoUl79}, the equivalence classes partition
the set of reachable states in $Q$.  Hence, for each equivalence class
$[\qp_i]$ there is a corresponding subset of $Q$.  Let $Q_1$ and $Q_2$
denote the subsets of $Q$ corresponding to the equivalence classes
$[\qp_1]$ and $[\qp_2]$ and assign an arbitrary order on each subset.
Select the first state, say $p_1 \in Q_1$, and define the set $R = \{q
\in Q_2\ |\ \exists n,m \in \Zf^+,\ \delta(p_1,x^m) = \delta(q,x^n) =
q\}$.  If there exists a state~$r \in R$ and string~$y \in \Sigma^+$
such that $\delta(r,y) = p_1$, then $M$ does not satisfy the partial
order condition, and this is a contradiction.  Otherwise, there does
not exist a $y \in \Sigma^+$ such that $\delta(r,y) = p_1$ for all $r
\in R$.  In this case there is a partial order on $p_1$ and on $Q_1
\backslash \{p_1\}$ because $p_1$ will never be visited again if $M$
reads a sufficient number of $x$s.  Remove $p_1$ from $Q_1$ and repeat
the procedure on $p_2 \in Q_1$.  After a finite number of iterations we
will either find a $p_i$ that satisfies our requirements, which means
that $M$ does not satisfy the partial order condition and is a
contradiction, or none of the states in $Q_1$ will have the required
characteristics, in which case $\Mp$ satisfies the partial order
condition.  Therefore, if $M$ satisfies the partial order condition, so
will its minimal equivalent $\Mp$.
\end{proof}}

\begin{lemma}\label{inter_po} 
Let $\Lp$ and $\Lpp$ be languages that satisfy the partial order
condition.  Then $L = \Lp \cap \Lpp$ also satisfies the partial order 
condition.
\end{lemma}
\SaveProof{\leminterpoproof}{lemma_inter_po}{\begin{proof}
Let $\Mp = (\Qp,\Sigma,\deltap,\qp_0,\Fp)$ be the minimal DFA accepting
the language $\Lp$ and let $\Mpp = (\Qpp,\Sigma,\deltapp,\qpp_0,\Fpp)$
be the minimal DFA accepting the language $\Lpp$.  We first construct
an automaton $M$ that accepts $\Lp \cap \Lpp$ by combining $\Mp$ and
$\Mpp$ using a direct product.  Define $M = (Q,\Sigma,\delta,q_{00},F)$
where $Q = \Qp \times \Qpp$, $q_{00} = (\qp_0,\qpp_0)$, $F = \{
(\qp,\qpp) \in Q\ |\ \qp \in \Fp\ \wedge\ \qpp \in \Fpp\}$ and
$\delta((\qp,\qpp),\sigma) = (\deltap(\qp,\sigma),
\deltapp(\qpp,\sigma))$.

We argue that if $\Mp$ and $\Mpp$ satisfy the partial order condition,
then so will $M$.  Assume, by contradiction, that $M$ does not satisfy
the partial order condition.  Then there exist two states $q_{ij} =
(\qp_i,\qpp_j)$ and $q_{kl} = (\qp_k,\qpp_l)$ and strings $x,y,z \in
\Sigma^+$ such that $\delta(q_{ij},x) = \delta(q_{kl},x) = q_{kl}$,
$\delta(q_{kl},y) = q_{ij}$ and $\delta(q_{ij},z) \in F$ if and only if
$\delta(q_{kl},z) \not\in F$.  In the first case assume that either $i
\not= k$ or $j \not= l$, and without loss of generality, assume the
former.  Then there exists state~$\qp_i \in \Qp$ and state~$\qp_k \in
\Qp$ such that $\deltap(\qp_i,x) = \deltap(\qp_k,x) = \qp_k$,
$\delta_1(\qp_k,y) = \qp_i$.  But this means that $\Mp$ does not
satisfy the partial order condition, a contradiction.  In the second
case assume that $i = k$ and $j = l$.  This implies that $q_{ij} =
q_{kl}$ and hence there cannot exist a string $z$ that distinguishes
the two states, also a contradiction.  Therefore $M$ must satisfy the
condition.

Since $M$ satisfies the partial order condition and accepts $L$,
by Lemma~\ref{po_min} the minimal automaton that accepts $L$ satisfies 
the partial order condition, and hence $L$ itself, satisfies the partial 
order condition.
\end{proof}}

\begin{theorem}\label{po_int}
If the partial order condition is sufficient for acceptance with bounded
error by MM-QFAs then the class $\RMM$ is closed under intersection.
\end{theorem}
\begin{proof}
By Lemma~\ref{inter_po} the intersection of two languages that satisfy
the partial order condition is a language that satisfies the partial
order condition.
\end{proof}

One method for proving that the class $\RMM$ is not closed under
intersection involves intersecting two languages in $\RMM$ and showing
that the resulting language is not in $\RMM$.  By Theorem~\ref{po_int}
this method will not work unless the partial order condition is
insufficient.  To study whether the partial order condition is
sufficient, as well as necessary, we show that a well known class of
languages can be accepted by an MM-QFA with bounded error.

\subsection{Piecewise Testable Sets}
A piecewise testable set is a boolean combination of sets of the form
\[L_z = \Sigma^*z_1\Sigma^*z_2\Sigma^*...\Sigma^*z_n\Sigma^*\] where
$z_i \in \Sigma$~\cite{HoTCS}.  Intuitively, $L_z$ is the language of
strings that contain the successive symbols of $z$ as a subsequence; we
call such a language a partial piecewise testable set.

Piecewise testable sets, introduced by Simon in~\cite{Si75}, form a
natural family of star-free languages.  Such sets define a class of
computations that wait for a partially ordered sequence of trigger
events (input symbols); if a trigger event (symbol) is read that is not
next in the sequence, it is simply ignored.  Another natural
interpretation of piecewise testable sets is subsequence searching.
Consider a language where a word is said to be in the language if it
contains a finite boolean combination of subsequences.  Such a language
is a piecewise testable set and word acceptance corresponds to
searching the words for the required subsequences.  Finally, such
languages belong to a class of languages whose MM-QFAs have an
arbitrarily large, but finite, set of ordered states.

We show, by construction, that MM-QFAs can accept partial piecewise
testable sets with bounded error.  The MM-QFAs we construct accept with
one-sided error and are what we call `end-decisive'.  We say that an
MM-QFA accepts with positive one-sided error if it accepts strings in
the language with non-zero probability and rejects strings not in the
language with certainty.  We say that an MM-QFA accepts with negative
one-sided error if it accepts strings in the language with certainty
and rejects strings not in the language with non-zero probability.

We say that an MM-QFA is end-decisive if it will not be observed
in an accept state until the end-marker $\$$ is read.  An MM-QFA
is co-end-decisive, if it will not be observed in a reject state
until the end-marker is read.  

Classes of languages that are accepted by end-decisive MM-QFAs with the
same one-sided error, i.e., all positive or all negative, are closed
under intersection and union.  Furthermore, if language $L$ can be
accepted by an end-decisive MM-QFA with bounded error, and language
$\Lp$ can be accepted by an end-decisive MM-QFA with bounded one-sided
error, then the union or intersection of $L$ and $\Lp$ can be accepted
by an end-decisive MM-QFA with bounded error.  To construct these
MM-QFAs we introduce two useful concepts:  junk states and trigger
chains.

A junk state is a halting state of an end-decisive or co-end-decisive
MM-QFA.  If the MM-QFA is end-decisive, then all its junk states are
reject states.  If the MM-QFA is co-end-decisive, then all its junk
states are accept states.  An end-decisive or co-end-decisive MM-QFA
may be observed in a junk state at any point of the computation.  While
junk states are either accept or reject states, we treat the junk state
as a separate halting state.  Any accept or reject state that is not a
junk state is called a decisive state.  Intuitively, a junk state
signals a failed computation.

Each, end-decisive or co-end-decisive automata that accepts with
bounded error has probability, bounded by some constant $\tau < 1$ of
ending up in a junk state and a probability $1-\tau$ of ending up in a
decisive state.  If $\tau \not< 1$ then the amount of probability
amplitude ending up in a decisive state can become arbitrary small,
dropping below any fixed margin. Thus, $\tau$ must be strictly less
than one for the MM-QFA to accept with bounded error; $\tau$ is
independent of the input string~$x$.

A trigger chain is a construction of junk states and transition
matrices that causes a reduction in amplitude of a particular state
only if the amplitude of another state is decreased, presumably by some
previous transition.  Trigger chains correspond directly to
partial piecewise testable sets.  Consider the matrix
\[X = \left[\begin{array}{ccc}
               \Half   &  \RtHalf &  \Half \\
               \RtHalf & 0        & -\RtHalf \\
               \Half   & -\RtHalf &  \Half \\
            \end{array}\right].\]
This matrix is a special case of a transition matrix introduced by
Ambainis and Freivalds\cite{AmFr98}.  This matrix operates on three
states and is a triggering mechanism of the chain.  Consider the vectors
\[\Ket{\psi} = (\alpha,0,\beta)^T\]
and
\[X\Ket{\psi} = \left(\frac{\alpha}{2} + \frac{\beta}{2},
                 \frac{\alpha}{\sqrt{2}} - \frac{\beta}{\sqrt{2}},
                 \frac{\alpha}{2} + \frac{\beta}{2}\right)^T.\]
The vectors $\Ket{\psi}$ and $X\Ket{\psi}$ are equal if and only if
$\alpha = \beta$.  If $\alpha \not= \beta$ then the
amplitudes of the first and third state are averaged, with the
remainder of the amplitude going into the second state.  We define a
generalized version of $X$ by embedding it into a larger identity block
matrix.  Define $X_i$ to be
\[X_i = \left[\begin{array}{ccc}
              I_i &   &               \\
                  & X &               \\
                  &   & I_{s - i - 3} \\
        \end{array}\right]\]
where $I_m$ is an $m\times m$ identity matrix, $X$ is defined as above,
and $s$ is the number of states, i.e. the size of $X_i$.  The matrix
$X_i$ operates on a triple of states, $q_i$ through to $q_{i+2}$.  We
assume that state $q_{i+1}$, the second state, is a junk state unless
otherwise noted.  

\begin{theorem}\label{thm:partial}
Let $L_z$ be a partial piecewise testable set.  There exists an
end-decisive MM-QFA that accepts $L_z$ with bounded positive one-sided
error.
\end{theorem}
\begin{proof}
We construct an MM-QFA $M$ with $m+1$ states that accepts $L_z$ where 
$z = z_0z_1...z_n$ and $m = 2n + 4$.

For each link in the trigger chain we require a junk state and a
non-halting state.  We order the states to correspond with the
description of the $X_i$ matrices.  Specifically, the first $2n+2$
states are the non-halting states, interleaved with junk states.  Each
triple of states $(q_{2i},q_{2i+1},q_{2i+2})$ corresponds to a link of
the trigger chain, of which there are $n+1$.  State~$q_{2n+1}$ is the
decisive accept state and state~$q_{2n+3}$ is the decisive reject
state.  The junk states are rejecting states.

Let $m = 2n + 4$ and $M = (Q,\Sigma,\delta,q_0,\Qacc,\Qrej)$ where 
\begin{eqnarray*}
Q      & = & \{q_0,...,q_m\} \\
\Qjunk & = & \{q_i \in Q\ |\ 0 < i < 2n\ \wedge\ i \equiv 1 \bmod{2}\} \cup
             \DefSep \{q_{2n+4}\} \\
\Qacc  & = & \{q_{2n+1}\} \\
\Qrej  & = & \{q_{2n+3}\}.
\end{eqnarray*}

Define $\delta$ by the transition matrices $\{U_\sigma\}_{\sigma \in
\Sigma}$.  Each transition matrix $U_\sigma$ consists of a product of
matrices:
\[U_\sigma = U_{\sigma,0}U_{\sigma,1}...U_{\sigma,n}\]
where the matrices $U_{\sigma,i}$ implement the triggers.  

Define $U_{\sigma,i}$ to be
\[U_{\sigma,i} = \left\{\begin{array}{lll}
                     S        & \  & i = 0\ \wedge\ z_0 = \sigma \\
                     X_{2i-2} & \  & 1 \leq i \leq n\ \wedge\ z_i = \sigma \\
                     I_{m+1}  & \  & \mathrm{otherwise} \\
                     \end{array}\right.\]
where
\[S = \left[\begin{array}{ccc}
             0 & 1 &         \\
             1 & 0 &         \\
               &   & I_{m-1} \\
            \end{array}\right]. \]
The matrix $S$ shifts the amplitude of $q_0$ to the junk state $q_1$.
This is the first trigger that is activated when $z_0$ is read.  

Finally, 
let the transition matrix for the end-marker $\$$ be
\[U_{\$} = FX_{2n}\]
where
\[F = \left[\begin{array}{cccccccc}
              R &        &   &   &   &   &   &   \\
                & \ddots &   &   &   &   &   &   \\
                &        & R &   &   &   &   &   \\
                &        &   & 0 & 0 & 0 & 0 & 1 \\ 
                &        &   & 0 & 1 & 0 & 0 & 0 \\ 
                &        &   & 0 & 0 & 0 & 1 & 0 \\ 
                &        &   & 0 & 0 & 1 & 0 & 0 \\ 
                &        &   & 1 & 0 & 0 & 0 & 0 \\ 
              \end{array}\right]\]
and the matrix 
\[R = \left[\begin{array}{cc}
            0 & 1 \\
            1 & 0 \\
            \end{array}\right].\]
The matrix $F$ sends all amplitude into the junk states.  The matrix
$X_{2n}$ sends some minimum amount of amplitude into an accept state if
the amplitudes of states $q_{2n}$ and $q_{2n+2}$ differ.

The initial configuration of the machine is $\Ket{\psi_{init}} =
(\alpha_0,\alpha_1,...,\alpha_m)^T$ where
\[\alpha_i = \left\{\begin{array}{lr}
     \frac{1}{\sqrt{n+2}} & 0 \leq i \leq 2n+2\  \wedge\  i\equiv 0\pmod{2}\\
     0                    & \mathrm{otherwise} \\
             \end{array}\right.\]
i.e. the amplitude is evenly distributed among all
non-halting states.  

The only decisive accepting state in the machine is $q_{2n + 1}$, and
amplitude only flows into it when the end-marker is read.  In order for
it to get a non-zero amplitude, the amplitudes of states $q_{2n}$ and
$q_{2n+2}$ must differ.  Since all non-halting states start with the
same amplitude, and since the amplitude of state~$q_{2n+2}$ will not
change during the execution of the machine until the end-marker is
read, the amplitude of state~$q_{2n-2}$ must change in order for the
amplitude of state~$q_{2n}$ to change.  Following the same argument,
state $q_{2i}$ will not change in amplitude, until state $q_{2i-2}$
changes in amplitude.  Furthermore, the change in amplitude of state
$q_{2i}$ is governed by the matrix components $X_{2i-2}$ and $X_{2i}$.
Hence, the initial change of amplitude of state~$q_{2i}$ depends
exclusively on a change in amplitude of state~$q_{2i-2}$ and is
governed by component $X_{2i-2}$ that is located in the transition
matrix $U_{z_i}$.  If any other transition matrix is applied, then the
amplitude of state~$q_{2i}$ will not change.  Hence, $M$ can read
$(\Sigma - \{z_i\})^*$ without changing the amplitude of
state~$q_{2i}$, but, as soon as $z_i$ is read, component $X_{2i-2}$
will be applied and $q_{2i}$ will have a decreased amplitude, provided
state~$q_{2i-2}$ already had a decrease of its amplitude.  Finally, the
amplitude of any state $q_{2i}$ will never increase beyond its initial
value, and once the amplitude of state~$q_{2i}$ decreases, it will
never increase beyond $\frac{1}{\sqrt{n+2}}(1 - (\frac{1}{2})^{n+1})$.
For the case of symbol~$z_0$, the amplitude of state~$q_0$ is changed
by matrix $S$ to $0$ and is the starting trigger.  When the end-marker
is read a minimum of $\frac{1}{\sqrt{2(n+2})}(\frac{1}{2})^{n+1}$ of
amplitude is placed into the accepting state only if the amplitude of
state $q_{2n}$ has decreased.  The amplitude from $q_{2n+2}$ is
channeled into a decisive reject state.  The rest of the amplitude,
from the remaining $n+1$ non-halting states is channeled into junk
states.  If the amplitudes of $q_{2n}$ and $q_{2n+2}$ do not differ
then all amplitude is channeled into junk and decisive reject states.

The probability of $M$ accepting a string not in the language is $0$,
while the probability of $M$ accepting a string in the language is at
least $\frac{1}{n+2}(\Half)^{2n+3}$.  We select the cut-point to be 
strictly between the two values.
\end{proof}

Any boolean combination of partial piecewise testable sets may be
expressed as a union of intersections of partial piecewise testable
sets and complements of partial piecewise testable sets, i.e.,
\begin{equation}
\bigcup^s_{i=0} \left(\bigcap^t_{j=0} \Ltilde_{ij}\right)
\end{equation}
where $\Ltilde_{ij}$ is a partial piecewise testable set or the 
complement thereof.

We first show how to construct the implicants of the above expression,
i.e. $\cap^t_{j=0} \Ltilde_{ij}$, and then, how to take the union of the 
implicants.  An implicant can be written in the form 
\[\bigcap^t_{j=0} \Ltilde_{ij} = \left(\bigcap^r_{j=0} L_{ij} \right) \bigcap 
                              \left(\bigcap^t_{j=r} \Lbar_{ij} \right),\]
where the $L_{ij}$s are partial piecewise testable sets.  By De
Morgan's rule, the latter part of this expression can be rewritten as
$\overline{\cup^t_{j=r} L_{ij}}$.  Let $\Lcap_i = \cap^r_{j=0} L_{ij}$,
let $\Lcup_i = \cup^t_{j=r} L_{ij}$, and let $L_i = \Lcap_i \cap
\overline{\Lcup_i}$.

First, we show that $\Lcap_i$ can be accepted by an end-decisive MM-QFA
with bounded positive one-sided error.  Second, we show that
$\overline{\Lcup_i}$ can be accepted by an end-decisive MM-QFA with
bounded error.  Third, we show that $L_i$ can be accepted by an
end-decisive MM-QFA with bounded error.  Finally, we show that
$\cup^s_{i=0} L_i$ can be accepted by an end-decisive MM-QFA with
bounded error.  We first need two composition lemmas.

We say that an MM-QFA $M$ accepts $L$ with cut-point $\lambda$ and
maximum margin $\eta$ if for all $x \in \Sigma^*$,
\[\lambda - \eta < \Paccept{M(x)} < \lambda + \eta.\]
Usually, the maximum margin will be exponentially greater than the
margin $\epsilon$; this creates problems when we compose automata.

\begin{lemma} \label{lem:mmqfa_tensor} 
Let $M$ and $\Mp$ be end-decisive MM-QFAs that accept $L$ and 
$\Lp$ respectively, with cut-points $\lambda$ and $\lambdap$, margins
$\epsilon$ and $\epsilonp$, and maximum margins $\eta$ and $\etap$.
There exists an end-decisive MM-QFA $\Mpp$ such that the inequalities
\begin{eqnarray}
& (\lambda + \epsilon)\cdot(\lambdap + \epsilonp) \leq \Paccept{\Mpp(x)} 
  \leq (\lambda + \eta)\cdot(\lambdap + \etap) 
  & \forall x \in L \cap \Lp, \label{eqn:tensor_1} \\
& (\lambda - \eta)\cdot(\lambdap + \epsilonp) \leq \Paccept{\Mpp(x)} 
  \leq (\lambda - \epsilon)\cdot(\lambdap + \etap) 
  & \forall x \in \Lbar \cap \Lp, \label{eqn:tensor_2} \\
& (\lambda + \epsilon)\cdot(\lambdap - \etap) \leq \Paccept{\Mpp(x)} 
  \leq (\lambda + \eta)\cdot(\lambdap - \epsilonp) 
  & \forall x \in L \cap \Lpbar, \label{eqn:tensor_3} \\
& (\lambda - \eta)\cdot(\lambdap - \etap) \leq \Paccept{\Mpp(x)} 
  \leq (\lambda - \epsilon)\cdot(\lambdap - \epsilonp) 
  & \forall x \in \Lbar \cap \Lpbar \label{eqn:tensor_4}
\end{eqnarray}
are satisfied.
\end{lemma}
\begin{proof}
Let $M = (Q,\Sigma,\delta,q_0,\Qacc,\Qrej)$ and $\Mp = 
(\Qp,\Sigma,\deltap,\qp_0,\Qaccp,\Qrejp)$ be end-decisive MM-QFAs
that accept $L$ and $\Lp$.  Using these two MM-QFAs we construct an
MM-QFA $\Mpp = (\Qpp,\Sigma,\deltapp,\qpp_0,\Qaccpp,\Qrejpp)$ that
satisfies the above inequalities.

Let $\Qpp = Q \times \Qp$ and $\qpp_0 = (q_0,\qp_0)$.  The sets
of halting states are defined as
\begin{eqnarray*}
\Qaccpp  & = & \{(q_i,\qp_j) \in \Qpp\ |\ q_i \in \Qacc\ \wedge\ \qp_j \in
                \Qaccp\}, \\
\Qrejpp  & = & \{(q_i,\qp_j) \in \Qpp\ |\ (q_i \in \Qrej\ \vee\ \qp_j \in 
                \Qrejp),\}
\end{eqnarray*}
 and the transition function $\deltapp$ is defined as
\[\deltapp((q,\qp),\sigma,(r,\rp)) = \delta(q,\sigma,r) \cdot
                                     \deltap(\qp,\sigma,\rp),\] 
which is a tensor product of the transition functions $\delta$ and 
$\deltap$.

Since $M$ and $\Mp$ are end-decisive, i.e., the accepting states will
only have non-zero amplitude when the end-marker is read; thus the
MM-QFA $\Mpp$ will be end-decisive.

By the tensor product construction, the probability of $\Mpp$ accepting
$x$ is
\[\Paccept{\Mpp(x)} = \Paccept{M(x)} \cdot \Paccept{\Mp(x)}.\]
Since 
\begin{eqnarray*}
  & \lambda + \epsilon \leq \Paccept{M(x)} \leq \lambda + \eta 
    & \forall x \in L, \\
  & \lambda - \eta \leq \Paccept{M(x)} \leq \lambda - \epsilon 
    & \forall x \not\in L, \\
  & \lambdap + \epsilonp \leq \Paccept{\Mp(x)} \leq \lambdap + \etap 
    & \forall x \in \Lp, \\
  & \lambdap - \etap \leq \Paccept{\Mp(x)} \leq \lambdap - \epsilonp 
    & \forall x \not\in \Lp,
\end{eqnarray*}
multiplying out the probabilities yields the inequalities 
\ref{eqn:tensor_1}, \ref{eqn:tensor_2}, \ref{eqn:tensor_3}, and
\ref{eqn:tensor_4}.
\end{proof}

\begin{corollary} \label{cor:pos_intersec}
Let $M$ and $\Mp$ be end-decisive MM-QFAs that accept $L$ and 
$\Lp$ respectively, with bounded positive one-sided error.
There exists an end-decisive MM-QFA that accepts $L \cap \Lp$ with
bounded positive one-sided error.
\end{corollary}
\begin{proof}
Let $\lambda$, $\lambdap$, $\epsilon$, and $\epsilonp$ be the respective
cut-points and margins of MM-QFAs $M$ and $\Mp$.  Since 
$\lambda - \epsilon = \lambdap - \epsilonp = 0$, $\lambda + \epsilon >
0$, and $\lambdap + \epsilonp > 0$, the result follows from 
Lemma~\ref{lem:mmqfa_tensor}.
\end{proof}

We mentioned before that the maximum maximum margin of a MM-QFA that
accepts language $L$ could be exponentially greater than the margin.
This prevents us from directly constructing intersections or unions of
languages that are accepted by end-decisive MM-QFAs with bounded
error.  To get around this problem we use a tensor power technique to
magnify the ratio of the probability of a true positive to the probability
of a false positive.

\begin{lemma}\label{lem:power}
Let $M$ be an end-decisive MM-QFA that accepts words in $L$ with
probability at least $\lambda + \epsilon$ and accepts words not in $L$
with probability at most $\lambda - \epsilon$.  For any positive
integer $n$ there exists an MM-QFA $\Mp$ that accepts words in $L$ with
probability at least $(\lambda + \epsilon)^n$, and accepts words not in
$L$ with probability at most $(\lambda - \epsilon)^n$.
\end{lemma}
\begin{proof}
Using Lemma~\ref{lem:mmqfa_tensor} to compose $n$ copies of $M$ 
yields the result.
\end{proof}

We first use Lemma~\ref{lem:power} to construct finite unions of
languages that are accepted by end-decisive MM-QFAs with bounded error.

\begin{lemma}\label{lem:union}
Let $M$ be an MM-QFA that accepts $L$ with bounded error and let $\Mp$
be an MM-QFA that accept $\Lp$ with bounded error.  There exists an 
MM-QFA $\Mpp$ that accepts $\Lpp = L \cup \Lp$ with bounded error.
\end{lemma}
\begin{proof}
Assume that $M$ accepts words in $L$ with probability at least $\lambda
+\epsilon$ and accepts words not in $L$ with probability at most
$\lambda - \epsilon$.  Similarly, assume that $\Mp$ accepts words in
$\Lp$ with probability at least $\lambdap +\epsilonp$ and accepts words
not in $\Lp$ with probability at most $\lambdap - \epsilonp$.

Using Lemma~\ref{lem:power} let $M_s$ be the $s$th tensor power of $M$
and $\Mp_t$ be the $t$th tensor power of $\Mp$.

Let $M_s = (Q,\Sigma,\delta,q_0,\Qacc,\Qrej)$ and $\Mp_t = 
(\Qp,\Sigma,\deltap,\qp_0,\Qaccp,\Qrejp)$, where $Q = \{q_0,...,q_{n-1}\}$ 
and $\Qp = \{\qp_0,...,\qp_{m-1}\}$.  Let $\delta$ and $\deltap$ be
represented by the unitary matrices $U_\sigma$ and $\Up_\sigma$
respectively.  

Let $\Mpp = (\Qpp,\Sigma,\deltapp,\Qaccpp,\Qrejpp)$ where
$\Qpp = \{\qpp_0,...,\qpp_{n+m-1}\}$, $\deltapp$ is represented by the
matrices
\[\Upp_\sigma = \left[\begin{array}{cc}
                      U_\sigma & 0  \\
                      0        & \Up_\sigma \\
                      \end{array}
                      \right],\]
$\Qaccpp = \{ \qpp_i \in \Qpp\ |\ q_i \in \Qacc\ \vee\ \qp_{i-n} \in
\Qaccp\}$, and $\Qrejpp = \{ \qpp_i \in \Qpp\ |\ q_i \in
\Qrej\ \vee\ \qp_{i-n} \in \Qrejp\}$.  The automata is initialized with
the amplitude evenly divided between the states $\qpp_0$ and $\qpp_n$,
i.e., each state contains $\RtHalf$ amplitude.  Intuitively, $M$ and
$\Mp$ run in parallel, not interacting unless one of the two crashes.
In that case the computation is over.

If $x \in L \cap \Lp$ then 
\[\Paccept{\Mpp(x)} \geq \frac{(\lambda + \epsilon)^s + 
                               (\lambdap + \epsilonp)^t}{2},\]
if $x \in L \cap \Lpbar$ then
\[\Paccept{\Mpp(x)} \geq \frac{(\lambda + \epsilon)^s}{2},\]
if $x \in \Lbar \cap \Lp$ then 
\[\Paccept{\Mpp(x)} \geq \frac{(\lambdap + \epsilonp)^t}{2},\]
and if $x \in \Lbar \cap \Lpbar$ then 
\[\Paccept{\Mpp(x)} \leq \frac{(\lambda - \epsilon)^s + 
                               (\lambdap - \epsilonp)^t}{2},\]
The last case corresponds to $x \not\in \Lpp$.  By setting $s$ and $t$
appropriately, we can ensure that 
\[(\lambda - \epsilon)^s + (\lambdap - \epsilonp)^t \ll 
  \min{\{(\lambda + \epsilon)^s, (\lambdap + \epsilonp)^t\}}.\]
Hence, the MM-QFA $\Mpp$ accepts $L \cup \Lp$ with bounded error. 
Furthermore, $\Mpp$ is end-decisive because both $M_s$ and $M_t$ are 
end-decisive. 
\end{proof}

\begin{corollary}\label{cor:pos_union}
Let $M$ and $\Mp$ be end-decisive MM-QFAs that accept $L$ and 
$\Lp$ respectively, with bounded positive one-sided error.
There exists an end-decisive MM-QFA that accepts $L \cup \Lp$ with
bounded positive one-sided error.
\end{corollary}
\begin{proof}
Since $\lambda - \epsilon = \lambdap - \epsilonp = 0$, the same 
argument as in Corollary~\ref{cor:pos_intersec} applies.
\end{proof}

One useful property of languages that are accepted by end-decisive
MM-QFAs with bounded positive one-sided error is that we can usually
construct end-decisive MM-QFAs that can accept the complement such
languages with bounded error.  We say that an end-decisive MM-QFA
accepts with positive amplitude, if the amplitude in it's accept states
is always non-negative.

\begin{lemma}\label{lem:complement}
Let $L$ be a language that is accepted by an end-decisive MM-QFA with
bounded positive one-sided error and positive amplitude.  There exists 
an end-decisive MM-QFA that accepts $\Lbar$ with bounded error.
\end{lemma}
\begin{proof}
Let $M = (Q,\Sigma,\delta,q_0,\Qacc,\Qrej)$ be an end-decisive MM-QFA
that accepts $L$ with bounded positive one-sided error.  Since $M$
rejects all strings not in $L$ with certainty, for every computation of
$M$ on $x \not\in L$ zero amplitude is placed into the accepting states
of $M$.  Let $n = |Q|$, let $a = |\Qacc|$ and assume that 
$\Qacc = \{q_{n-1},q_{n-2},...,q_{n-a}\}$.

We use $M$ to construct an end-decisive MM-QFA $\Mp$ to accept $\Lbar$
with bounded error.  Let $\Mp = (\Qp,\Sigma,\deltap,q_0,\Qaccp,\Qrejp)$
where 
\begin{eqnarray*}
\Qp    & = & Q \cup \{q_n,q_{n+1},...,q_{n+3a}\} \\
\Qrejp & = & \Qrej \cup \Qacc \cup \{ q_{n+i} \in \Qp\ |\ i\equiv2\bmod 3\}\\
\Qaccp & = & \{ q_{n+i} \in \Qp\ |\ i \equiv 0 \bmod 3\}
\end{eqnarray*}
and the transition function $\deltap$ is extended in the following
manner.   For all symbols except the end-marker, the transition function
for $\Mp$ is defined by the matrices
\[\Up_\sigma = \left[\begin{array}{cc}
                    U_\sigma &  \\
                             & I_{3a} \\
                    \end{array} \right].\]
The end-marker transition is defined by the matrix 
\[\Up_\$ = \left[\begin{array}{cc}
                    U_\$ &  \\
                         & I_{3a} \\
                    \end{array} \right] X,\]
where matrix $X$ performs an averaging and cleanup operation.   We
define $X$ in terms of $4\times4$ sub-matrices.  Every accept state
$q_{n-a+i} \in \Qacc$ in $M$ becomes a reject state in $\Mp$.
Additionally, for each such state, 3 additional states were added to
$\Mp$, $q_{n+3i}$, $q_{n+3i+1}$, and $q_{n+3i+2}$, these are an
accepting, a non-halting, and a rejecting state respectively.  The
matrix $X$ operates on the 4-tuples of states $(q_{n-a+i},q_{n+3i},
q_{n+3i+1},q_{n+3i+2})$.  Each operation is localized to the 4-tuple of
states and hence can be described by a $4\times4$ matrix $X_i$.  Assume
that the order of rows and columns of the matrix correspond to the
order in the 4-tuple.  Let
\[X_i = \overbrace{\left[\begin{array}{cccc}
                         1 &   &   &   \\
                           & 1 &   &   \\
                           &   & 0 & 1 \\
                           &   & 1 & 0 \\
              \end{array}\right]}^{\mathrm{cleanup}}
        \overbrace{\left[\begin{array}{cccc}
                         \Half    &  \RtHalf  & \Half   &   \\
                        -\RtHalf  & 0         & \RtHalf &   \\
                         \Half    & -\RtHalf  & \Half   &   \\
                        &           &         & 1 \\
                        \end{array}\right]}^{averaging}.\]
Since $M$ accepts with positive amplitude, the amplitude in state
$q_{n-a+i}$ will be non-negative.  If the non-halting
state~$q_{n+3i+1}$ contains a fixed amount of amplitude $\alpha$, and
the old accept state~$q_{n-a+i}$ contains $\beta$ amplitude.  Then, the
averaging operation places $\frac{\alpha-\beta}{\sqrt{2}}$ amplitude in
the accept state~$q_{n+3i}$.  Then, the cleanup operation places any
amplitude remaining in the non-halting state~$q_{n+3i+1}$ into the
reject state~$q_{n+3i+2}$.

We initialize $\Mp$ in the same way as $M$ except that a fraction of the
amplitude is placed in the new non-halting states.  These states behave
as reservoirs until the end-marker is read.  The amount of amplitude
placed in the states is greater than the maximum amount of amplitude
that any accepting state may ever contain.

If $x \in L$ then at least one of the accept states of $M$ will contain
a minimum amount of positive amplitude.  Hence, the amount of amplitude
in at least one of the accept states of $\Mp$ will be strictly less than
$\frac{\alpha}{\sqrt{2}}$, by some fixed amount.  If $x \not\in L$ then
all accept states of $\Mp$ will have exactly $\frac{\alpha}{\sqrt{2}}$
amplitude in them.

Hence, if $x \in L$ the probability of $\Mp$ accepting $x$ will be 
strictly less than if $x \not\in L$.  Hence, $\Mp$ accepts $\Lbar$ with
bounded error.  Since the accept states are only observed after the 
end-marker is read, $\Mp$ is end-decisive.
\end{proof}

If $L$ is a language that can be accepted by an end-decisive MM-QFA 
with bounded error and $\Lp$ is a language that can be accepted by an
end-decisive MM-QFA with bounded positive one-sided error, then
we use Lemma~\ref{lem:power} to construct an MM-QFA that accepts the
intersection of the two languages.

\begin{lemma}\label{lem:intersection}
Let $M$ be an end-decisive MM-QFA that accepts $L$ with bounded error
and let $\Mp$ be an end-decisive MM-QFA that accept $\Lp$ with bounded 
positive one-sided error.  There exists an MM-QFA $\Mpp$ that accepts 
$\Lpp = L \cap \Lp$ with bounded error.
\end{lemma}
\begin{proof}
Let MM-QFA $M$ accept $L$ with cut-point $\lambda$, margin
$\epsilon$, and maximum margin $\eta$, and let MM-QFA $\Mp$ accept
$\Mp$ with cut-point $\lambdap$, margin $\epsilonp$, and maximum
margin $\etap$.

First, consider the inequalities in Lemma~\ref{lem:mmqfa_tensor} that
occur when we compose the MM-QFAs $M$ and $\Mp$ using the tensor 
technique.  Since MM-QFA $\Mp$ accepts with bounded positive one-sided
error, the inequalities are:
\begin{eqnarray*}
& (\lambda + \epsilon)\cdot(\lambdap + \epsilonp) \leq \Paccept{N(x)} 
  \leq (\lambda + \eta)\cdot(\lambdap + \etap) 
  & \forall x \in L \cap \Lp, \\
& (\lambda - \eta)\cdot(\lambdap + \epsilonp) \leq \Paccept{N(x)} 
  \leq (\lambda - \epsilon)\cdot(\lambdap + \etap) 
  & \forall x \in \Lbar \cap \Lp, \\
& (\lambda + \epsilon)\cdot 0 \leq \Paccept{N(x)} 
  \leq (\lambda + \eta)\cdot 0
  & \forall x \in L \cap \Lpbar, \\
& (\lambda - \eta)\cdot 0 \leq \Paccept{N(x)} 
  \leq (\lambda - \epsilon)\cdot 0
  & \forall x \in \Lbar \cap \Lpbar.
\end{eqnarray*}
These reduce to three cases:
\begin{eqnarray}
& \Paccept{\Mpp(x)} \geq (\lambda + \epsilon)\cdot(\lambdap + \epsilonp) 
  & \forall x \in L \cap \Lp, \\
& \Paccept{\Mpp(x)} \leq (\lambda - \epsilon)\cdot(\lambdap + \etap) 
  & \forall x \in \Lbar \cap \Lp, \\
& \Paccept{\Mpp(x)} = 0
  & \forall x \in \Lpbar.
\end{eqnarray}
If we can guarantee that 
\[(\lambda - \epsilon)\cdot(\lambdap + \etap) < 
  (\lambda + \epsilon)\cdot(\lambdap + \epsilonp)\]
then the tensor technique is sufficient to construct the intersection.
Let $M_n$ be the $n$th tensor composition of $M$.  By Lemma~\ref{lem:power}
$M_n$ accepts words in $L$ with probability at least $(\lambda +
\epsilon)^n$ and accepts words not in $L$ with probability at most
$(\lambda - \epsilon)^n$.  Construct MM-QFA $\Mpp$ by composing $M_n$ 
with $\Mp$ using the tensor technique; for sufficiently large constant $n$ 
the inequality 
\[(\lambda - \epsilon)^n\cdot(\lambdap + \etap) < 
  (\lambda + \epsilon)^n\cdot(\lambdap + \epsilonp)\]
will be satisfied.  Thus, MM-QFA $\Mpp$ accepts $\Lpp$ end-decisively 
with bounded error.
\end{proof}

We are now assemble our array of tools to construct an arbitrary boolean
combination of partial piecewise testable sets.

\begin{theorem}
Piecewise testable sets can be accepted by end-decisive MM-QFAs
with bounded error.
\end{theorem}
\begin{proof}
Let $L$ be a piecewise testable set.  We first rewrite it in canonical
form:
\begin{eqnarray*}
L & = & \bigcup^s_{i=0}\bigcap^t_{j=0}\Ltilde_{ij} \\
  & = & \bigcup^s_{i=0}\left( \cap^r_{j=0} L_{ij} \bigcap 
                              \cap^t_{j=r} \Lbar_{ij} \right) \\
  & = & \bigcup^s_{i=0}\left( \cap^r_{j=0} L_{ij} \bigcap 
                    \overline{\cup^t_{j=r} L_{ij}}    \right) \\
  & = & \bigcup^s_{i=0}\left( \Lcap_i \bigcap \overline{\Lcup_i} \right) \\
  & = & \bigcup^s_{i=0} L_i
\end{eqnarray*}

By Theorem~\ref{thm:partial} we can construct end-decisive MM-QFAs that
accept partial piecewise testable sets, $L_{ij}$ with bounded positive
one-sided error.  Using these constructions and
Corollaries~\ref{cor:pos_intersec} and~\ref{cor:pos_union}, we can
construct end-decisive MM-QFAs that accept languages $\Lcap_i$ and
$\Lcup_i$ with bounded positive one-sided error.

The constructions in Theorem~\ref{thm:partial} only channel
non-negative amplitude into their accept states, furthermore, the
constructions in Lemmas~\ref{lem:mmqfa_tensor} and~\ref{lem:union} do
not negate amplitude.  Consequently, the constructions for languages
$\Lcap_i$ and $\Lcup_i$ only channel positive amplitude into their
accept states.  Hence, said constructions accept with positive
amplitude.  Since $\Lcup_i$ is also accepted with bounded positive
one-sided error, by Lemma~\ref{lem:complement}, we can construct an
end-decisive MM-QFA that can accept $\overline{\Lcup_i}$ with bounded
error.

Since $\Lcap_i$ is accepted by an end-decisive MM-QFA with bounded
positive one-sided error and $\overline{\Lcup_i}$ is accepted by an
end-decisive MM-QFA with bounded error, by Lemma~\ref{lem:intersection}
we can construct an end-decisive MM-QFA that accepts $L_i = \Lcap_i
\cap \overline{\Lcup_i}$ with bounded error.

Since the languages $L_i$ can be accepted by end-decisive MM-QFAs with
bounded error, by Lemma~\ref{lem:union}, we can construct an
end-decisive MM-QFA that accepts $L = \cup_i L_i$ with bounded error.
\end{proof}

\section{Conclusions}
We defined two models of 1-way quantum finite automata: the
measure-once model that performs one measurement at the end of the
computation, and the measure-many model that performs a measurement
after every transition.  The measure-many model is strictly more
powerful than the measure-once but is more difficult to characterize.

When restricted to accepting with bounded error, measure-once automata
can only accept group languages, while unrestricted measure-once
automata can accept irregular sets and in particular, can solve the
word problem on the free group.  Any language accepted by a MO-QFA can
also be accepted by a PFA, therefore PFAs can also solve the word
problem on the free group.  We also sketched an algorithm for
determining equivalence of two MO-QFAs.

The measure-many automaton is difficult to characterize.  We have shown
that the two classes of languages, those accepted with and without
bounded error, are closed under complement and inverse homomorphisms;
it is still an open question if these classes are closed under boolean
operations.  We defined the partial order condition for languages and
proved that it is a necessary condition for a language to be accepted
by an MM-QFA with bounded error.  We also showed that piecewise
testable sets can be accepted with bounded error by MM-QFAs, and in the
process detailed several novel construction techniques.

We do not know if the partial order condition is also a sufficient
condition for bounded acceptance.  If it is then the two classes of
languages accepted by an MM-QFA are closed under intersection.

\bibliography{qfachar}

\begin{thebibliography}{ANTSV99}

\bibitem[ABFK99]{AmBoFrKi99}
A.~Ambainis, R.~Bonner, R.~Freivalds, and A.~Kikusts.
\newblock Probabilities to accept languages by quantum finite automata.
\newblock In {\em Computation and Combinatorics}, volume 1627 of {\em Lecture
  Notes on Computer Science}, 1999.

\bibitem[AF98]{AmFr98}
A.~Ambainis and R.~Freivalds.
\newblock 1-way quantum finite automata: Strengths, weaknesses and
  generalizations.
\newblock In {\em Proceedings of the 39th Annual Symposium on Foundations of
  Computer Science}, pages 332--342, November 1998.

\bibitem[AI99]{AmIw99}
M.~Amano and K.~Iwama.
\newblock Undecideability of quantum finite automata.
\newblock In {\em Proceedings of the 31st Annual ACM Symposium on the Theory of
  Computing}, pages 368--375, 1999.

\bibitem[ANTSV99]{AmNaTaVa99}
A.~Ambainis, A.~Nayak, A.~Ta-Shma, and U.~Vazirani.
\newblock Dense quantum coding and a lower bound for 1-way quantum automata.
\newblock In {\em Proceedings of the 31st Annual ACM Symposium on the Theory of
  Computing}, pages 376--383, 1999.

\bibitem[BV97]{BeVa97}
E.~Bernstein and U.~Vazirani.
\newblock Quantum complexity theory.
\newblock {\em SIAM Journal of Computing}, pages 1411--1473, October 1997.

\bibitem[Eil76]{Ei76}
S.~Eilenberg.
\newblock {\em Automata, Languages and Machines}, volume~B.
\newblock Academic Press, New York, 1976.

\bibitem[HU79]{HoUl79}
J.~Hopcroft and J.~Ullman.
\newblock {\em Introduction to Automata Theory, Languages and Computation}.
\newblock Addison-Wesley Publishers, Reading, Massachusetts, 1979.

\bibitem[KW97]{KoWa97}
A.~Kondacs and J.~Watrous.
\newblock On the power of quantum finite state automata.
\newblock In {\em Proceedings of the 38th Annual Symposium on Foundations of
  Computer Science}, pages 66--75, 1997.

\bibitem[LZ77]{LiZa77}
R.~Lipton and Y.~Zalcstein.
\newblock Word problem solvable in logspace.
\newblock {\em Journal of the ACM}, 24(3):523--526, July 1977.

\bibitem[MC00]{MoCr98}
C.~Moore and J.~Crutchfield.
\newblock Quantum automata and quantum grammars.
\newblock {\em Theoretical Computer Science}, 237:275--306, 2000.

\bibitem[MT69]{MeTh69}
A.~Meyer and C.~Thompson.
\newblock Remarks on algebraic decomposition of automata.
\newblock {\em Mathematical Systems Theory}, 3(2):110--118, 1969.

\bibitem[Nay99]{Na99}
A.~Nayak.
\newblock Optimal lower bounds for quantum automata and random access codes.
\newblock In {\em Proceedings of the 40th Annual Symposium on Foundations of
  Computer Science}, 1999.

\bibitem[Ort87]{Or87}
J.~Ortega.
\newblock {\em Matrix Theory}.
\newblock Plenum Press, New York, New York, 1987.

\bibitem[Paz71]{Pa71}
A.~Paz.
\newblock {\em Introduction to Probabilistic Automata}.
\newblock Academic Press, New York, New York, 1971.

\bibitem[Per94]{HoTCS}
D.~Perrin.
\newblock Finite automata.
\newblock In J.~van Leeuwen, editor, {\em Handbook of Theoretical Computer
  Science}, volume~B, chapter~1. Elsevier Science Publisher, 1994.

\bibitem[Pin87]{Pi87}
J.~Pin.
\newblock On languages accepted by finite reversible automata.
\newblock In {\em Proceedings of the 14th International Colloquium on Automata,
  Languages and Programming}, volume 267 of {\em Lecture Notes on Computer
  Science}, pages 237--249, 1987.

\bibitem[Rab63]{Ra63}
M.~Rabin.
\newblock Probabilistic automata.
\newblock {\em Information and Control}, 6:230--245, 1963.

\bibitem[Sim75]{Si75}
I.~Simon.
\newblock Peicewise testable events.
\newblock In {\em Proc. of the 2nd GI Conf}, volume~33 of {\em Lecture Notes on
  Computer Science}, 1975.

\bibitem[Wag85]{Wa85}
S.~Wagon.
\newblock {\em The Banach-Tarski Paradox}.
\newblock Cambridge University Press, New York, New York, 1985.

\end{thebibliography}

\appendix

\section{End-Marker Theorems}\label{equiv_mmqfa}
\begin{theorem}\label{mo_em}
Let $M$ be an MO-QFA that has both left and right end-markers. 
There exists an MO-QFA $\Mp$ that uses only one end-marker and is
equivalent to $M$.
\end{theorem}
\begin{proof}
Let $M = (Q,\Sigma,\delta,q_0,F)$ be an MO-QFA with left and right
end-markers, effectively allowing $M$ to start in any possible
configuration.  Define $\Mp = (Q,\Sigma, \deltap,q_0,F)$ from $M$.  Let
$\delta$ be defined in terms of the transition matrices
$\{U_\sigma\}_{\sigma \in \Sigma}$.  We define $\deltap$ from $\delta$
in the following way: for every $\sigma \in \Sigma$ let
\[\Up_\sigma = U^{-1}_\cent U_\sigma U_\cent\]
and let
\[\Up_\$ = U_\$ U_\cent \]
Now consider what happens when $M$ and $\Mp$ read a string $x =
x_1...x_n$.  Since 
\begin{eqnarray*}
\Up(x\$) & = & \Up_\$\Up_{x_n}...\Up_{x_1} \\
        & = & U_\$ U_\cent U^{-1}_\cent U_{x_n}U_\cent ...
              U^{-1}_\cent U_{x_1} U_\cent \\
        & = & U_\$ U_{x_n}...U_{x_1}U_\cent \\
        & = & U(\cent x\$),
\end{eqnarray*}
the probability of $M$ accepting $x$ is equal to the probability of 
$\Mp$ accepting $x$.  Thus one end-marker on the right
suffices, and by symmetry one left end-marker would also suffice.
Therefore, an MO-QFA starting in configuration $\Ket{q_0}$ can simulate
an MO-QFA starting in any arbitrary configuration.
\end{proof}

\begin{theorem}
Let $M$ be an MM-QFA that has both left and right end-markers. 
There exists an MM-QFA $\Mp$ that uses only a right end-marker and
is equivalent to $M$.
\end{theorem}
\begin{proof}
Let $M = (Q,\Sigma,\delta,\Qacc,\Qrej)$ be an MM-QFA
that uses two end-markers and accepts $L$.  Assume without loss of
generality that
\begin{eqnarray*}
\Qnon & = & \{q_i \in Q\ |\ 0 \leq i < n_{non} \} \\
\Qacc & = & \{q_i \in Q\ |\ n_{non} \leq i < n_{acc}\} \\
\Qrej & = & \{q_i \in Q\ |\ n_{acc} \leq i < n_{rej} = n = |Q|\},
\end{eqnarray*}
which facilitates a simpler description of $\Mp$.  We construct $\Mp =
(\Qp,\Sigma,\deltap,\Qaccp,\Qrejp)$ that accepts $L$ with only the
right end-marker.  Let $\Qp = Q \cup \{q_n,q_{n+1},...,q_{2n -
n_{non}}\}$, $\Qaccp = \{q_{n + i - n_{non}} \in \Qp\ |\ q_i \in
\Qacc\}$ and $\Qrejp = \{q_{n + i - n_{non}} \in \Qp\ |\ q_i \in
\Qrej\}$.  Assume that $\delta$ is defined in terms of transition
matrices $\{U_\sigma\}_{\sigma \in \Sigma}$.  The construction of
$\{\Up_\sigma\}_{\sigma \in \Sigma}$ is similar to that in the proof of
Theorem~\ref{mo_em}.  Let $I_l$ represent an identity matrix of size
$l$ and $m = n - n_{non}$.  We define $\deltap$ in terms of its
unitary block matrices.  For all $\sigma \in \Sigma$ let
\begin{eqnarray*}
\Up_\sigma & = & \left[\begin{array}{cc}
                     U^{-1}_\cent &  \\
                                  & I_m \\
                     \end{array} \right]
                     S
                     \left[\begin{array}{cc}
                     U_\sigma &  \\
                              & I_m \\
                     \end{array} \right]
                     \left[\begin{array}{cc}
                     U_\cent &  \\
                             & I_m \\
                     \end{array} \right]\\
\Up_\$ & = & S   \left[\begin{array}{cc}
                     U_\$ &  \\
                              & I_m \\
                     \end{array} \right]
                     \left[\begin{array}{cc}
                     U_\cent &  \\
                             & I_m \\
                     \end{array} \right]
\end{eqnarray*}
where
\[S =            \left[\begin{array}{ccc}
                     I_{n_{non}} &     &     \\
                                 &     & I_m \\
                                 & I_m &     \\
                     \end{array} \right]\]
transfers (sweeps) all probability amplitude from states in the old
halting states to the new halting states.  The old halting states,
those in $\Qacc$ and $\Qrej$, are no longer halting states in $\Mp$.
The operation of $\Mp$ is similar to the operation of the QFA
constructed in Theorem~\ref{mo_em}, The ``sweeping'' operation saves the
amplitude that was in the old states, while it performs the
$U^{-1}_\cent$ operation in the new halting states (since otherwise the
$U^{-1}_\cent$ would corrupt the amplitude stored in the original halting
states).
\end{proof}

\AppendixProof{\lemmauidproof}{lemma_u_id}{Proof of Lemma~\ref{u_id}}
\AppendixProof{\theoremmoqfagfaproof}{theorem_moqfagfa}{Proof of
                                                       Theorem~\ref{moqfagfa}}
\AppendixProof{\theoremmmcloseproof}{theorem_mm_close}{Proof of 
                                                       Theorem~\ref{mm_close}}
\AppendixProof{\theorempocondproof}{theorem_po_cond}{Proof of 
                                                       Theorem~\ref{po_cond}}
\AppendixProof{\lemmapominproof}{lemma_po_min}{Proof of Lemma~\ref{po_min}}
\AppendixProof{\leminterpoproof}{lemma_inter_po}{Proof of Lemma~\ref{inter_po}}
\AppendixProof{\theorempartialproof}{theorem_partial}{Proof of 
                                                       Theorem~\ref{partial}}
\AppendixProof{\lemintersecproof}{lemma_intersec}{Proof of Lemma~\ref{intersec}}

\end{document}